\documentclass{aa}  

\usepackage{graphicx}
\usepackage{txfonts}
\usepackage[colorlinks=true,     linkcolor=blue, citecolor=blue, filecolor=blue, urlcolor=blue]{hyperref}

\usepackage{color}

\begin{document}

   \title{ Outflow from the very massive Wolf-Rayet binary Melnick 34 \thanks{Based on observations made with ESO telescopes
   at the Paranal observatory under programmes  60.A-9351, 0104.D-0084 and 0104.D-0084.}}

   \author{N. Castro\inst{1,2}, P. M. Weilbacher\inst{1}, M. M. Roth\inst{1}, P. A. Crowther\inst{3}, A. Monreal-Ibero\inst{4},
   J. Brinchmann\inst{5} \and G. Micheva\inst{1}
   }

   \institute{Leibniz-Institut für Astrophysik Potsdam (AIP), An der Sternwarte 16, 14482, Potsdam, Germany
                        \email{ncastro@aip.de}
                        \and
                        Institut für Astrophysik, Georg-August-Universität, Friedrich-Hund-Platz 1, 37077 Göttingen, Germany
                        \and
                        Department of Physics \& Astronomy, University of Sheffield, Hounsfield Road, Sheffield, S3 7RH, UK
                        \and
                        Leiden Observatory, Leiden University, P.O. Box 9513, 2300 RA Leiden, The Netherlands
                        \and
                        Instituto de Astrofísica e Ciências do Espaço, Universidade do Porto, CAUP, Rua das Estrelas, PT4150-762 Porto, Portugal
}

\authorrunning{N. Castro et al.}
\titlerunning{Outflows from Melnick 34}

\abstract{      Melnick 34 (Mk\,34) is one of the most massive binary systems known and is one of the brightest X-ray point sources in the 30 Doradus region. We investigated the impact of this massive system on the surrounding interstellar medium (ISM) using the optical spectroscopic capabilities of the narrow-field mode (NFM) of the Multi-Unit Spectroscopic Explorer (MUSE). MUSE-NFM spatially resolved the ISM in the vicinity of Mk\,34 with a resolution comparable to that of the HST. The analysis of the [N\textsc{ii}]\,$\lambda$\,6583 and [S\textsc{ii}]\,$\lambda$\,6717 emission lines reveals a cone-like structure apparently originating from Mk\,34 and extending southeast. Electron density maps and radial velocity measurements of the ISM lines further support an outflow scenario traced by these emissions. While no clear northwestern counterpart to this outflow was observed, we note increased extinction in that direction, towards the R136 cluster. The ISM material along the projected diagonal of the outflow on both sides of Mk\,34 shows similar properties in terms of the emission line ratios seen in the Baldwin-Phillips-Terlevich diagram. These results are consistent across two observational epochs. Additionally, we examined the residual maps within a 0.5" radius of Mk\,34 after modeling and subtracting the point spread function. The observed variations in the residuals could potentially be linked to Mk\,34's known periodic behavior. However, further observations with appropriate cadence are needed to fully monitor the 155\,day periodicity of Mk\,34's X-ray emissions.}

\maketitle % 
\nolinenumbers

%________________________________________________________________
\section{Introduction}

Very massive stars ($>100\,$M$_\odot$) are rare in the Universe \citep{2015HiA....16...51V}. However, massive OB stars are one of the main energy drivers and sources of processed chemical elements in the interstellar medium (ISM), and they chemically and dynamically shape the galaxies. Massive stars are also the most plausible sources of the ionization of the early Universe \citep{2022arXiv220201413E}. Their strong stellar winds and outflows can punch holes in the ISM, through which ionizing radiation leaks \citep{2018A&A...611A..95W,2020A&A...644A..21R}. The evolution of very massive stars and their interplay with the ISM must first be understood in our local neighborhood so that we may obtain a clearer view of the related phenomena at higher redshifts \citep{2017ApJ...845..165M} and in the early stages of the Universe.

The heart of the Tarantula nebula, NGC\,2070, in the Large Magellanic Cloud (LMC) hosts the most massive stars known that have been spatially resolved \citep{2010MNRAS.408..731C}. The distance to the LMC of 49.9\,kpc \citep{2013Natur.495...76P}  means that we can study them individually and together with the ISM, allowing us to explore the stellar feedback from very massive stars in the surrounding medium. Melnick 34 (also known as Mk\,34, BAT99 116, and Brey 84) is one of the most luminous  Wolf-Rayet (WR) stars in NGC\,2070 and is the main X-ray source \citep{2022MNRAS.515.4130C} after the core of the nebula itself, WR R136a1 \citep{2013A&A...558A.134D}.  Mk\,34 was initially classified as a single WN5h \citep{1998MNRAS.296..622C} and subsequently as a WN5h:a \citep{2008MNRAS.389..806S,2011MNRAS.416.1311C}. It was claimed to have an initial record mass of $390\,$M$_\odot$ \citep{2014A&A...565A..27H}. Further multi-epoch studies showed a more complex picture, wherein Mk\,34 is actually a highly eccentric binary of two very massive stars \citep{2011IAUS..272..497C}. The duplicity of Mk\,34 was confirmed by the strong X-ray emission, likely induced by the colliding winds  \citep{2006AJ....131.2164T}. Follow-up observations in X-rays revealed a periodic variability of 155\,days and an average luminosity that is one order of magnitude higher than that of any comparable WR binary in the Milky Way \citep{2018MNRAS.474.3228P} and $\sim$23\% higher than that of $\eta$\,Carinae. Additional optical parameters confirm that Mk\,34 is a spectroscopic binary of two similar WN5h stars \citep{2019MNRAS.484.2692T}, which may end its life as a double black hole system \citep{2022ApJ...925...69B}.

Mk\,34 is apparently in relative isolation from other bright stars \citep{2018AA...614A.147C}, and is far enough from the core of the cluster (10", 2.4\,pc) to be individually resolved. It is a perfect laboratory in which to explore the feedback mechanisms of very massive stars in the surrounding medium. We spectroscopically and spatially mapped Mk\,34 using the state-of-the-art integral field spectrograph of Multi-Unit Spectroscopic Explorer (MUSE; \citealt{2014Msngr.157...13B,2019AN....340..989R}) at the Very Large Telescope. MUSE in its narrow-field-mode configuration (NFM) allowed us to explore Mk\,34 and the surrounding ISM (field of view of 7.5"$\times$7.5", 1.8$\times$1.8\,pc) from a ground-based telescope but with a spatial resolution similar to that of the \textit{Hubble} Space Telescope \citep{2021Msngr.182...50C}.

In the present study, we report the detection of apparent  outflow signatures close to the Mk\,34 binary system and imprints in the ISM using high-spatial-resolution optical spectroscopic data obtained with MUSE. After detailing the data used in this work in Sect.~\ref{data}, we present the properties of the ISM surrounding Mk\,34 in Sect.~\ref{ISM}. In Sect.~\ref{epoch}, we compare the results obtained across two available epochs. We model the point spread function of MUSE-NFM in Sect.~\ref{psf} and examine the residual maps for evidence of possible outflows in the proximity of the binary. In Section~\ref{bpt} we explore the Baldwin-Phillips-Terlevich (BPT) diagram in the region around the Mk\,34 outflow, searching for varying ionizing sources within the ISM to help trace and confirm this phenomenon. Finally, in Sect.~\ref{conclusion}, we present a discussion of the findings of this study and outline potential future directions to continue this work.

\section{Data}
\label{data}

Mk\,34 MUSE-NFM data were obtained as part of the ESO observing programs 0104.D-0084 and 108.224L (PI N. Castro), which focused on spectroscopically resolving the core of NGC\,2070, the R136 cluster \citep{2021Msngr.182...50C}. The MUSE instrument \citep{2014Msngr.157...13B,2010SPIE.7735E..08B}, mounted on the Very Large Telescope in Chile, in its NFM configuration offers a field of view of 7.5"$\times$7.5" and a pixel scale of 0.025". The spectral information  in each pixel spans 4800\,--\,9300\,\AA, with a resolving power of R\,$\approx$\,3000  around H$\alpha$.

The field centered in Mk\,34 was first observed on 25 November 2019. Under the weather conditions of that night (photometric sky with ambient seeing between 0.57" and 0.83"), and with the performance of the adaptive optics, we achieved an image quality of $\approx$\,80\,mas. The field was observed for a total exposure time of 2520 seconds. The integration time was split into four 630\,s exposures, applying small spatial dithers (up to 0.1") between each of them. A second epoch was observed on 9 December 2021 in clear conditions (and ambient seeing of 0.4" to 0.58"), with identical exposure times, but this time with field rotations of 90$^\circ$ between each exposure to better suppress systematic errors. The combined cube of the Mk\,34 field delivers a central AO-corrected peak with an FWHM of $\approx$\,65\,mas in the $I$-band; however, the Strehl ratio is low and the overall point spread function is best described by a diffraction-limited Moffat-like core and a halo of uncorrected seeing \citep{2019A&A...628A..99F}.

The MUSE-NFM data were reduced with the  MUSE pipeline \citep{musepipeline} and included most of the standard processing steps. The basic processing, including CCD-level calibrations, such as bias correction, flat-fielding, wavelength calibration, geometrical calibration, illumination correction, and smoothed spatial twilight-flat correction, was carried out with v2.8 of the pipeline within the MuseWise \citep{musewise} environment. The resulting "pixel tables" were extracted from the database and the post-processing steps were then carried out with v2.8.1 called from EsoRex. In NFM, the atmospheric refraction is normally corrected by the hardware dispersion corrector, but due to the high zenith distances of the observations (airmass 1.4 to 1.5), significant residuals (up to 60\,mas) were visible. We therefore implemented an empirical correction as a patch to the standard MUSE pipeline, whereby we first reconstructed a cube coarsely sampled in 10\,\AA\ wavelength steps, measured the centroid for a selected star in each of them, and computed a fourth-order polynomial for both spatial directions. This polynomial was applied to the coordinates in the pixel table of the exposure in reverse in order to correct the residual refraction. This extra procedure resulted in a correction to below the spatial MUSE sampling ($\lesssim 25$\,mas). We continued to follow the standard processing steps: The data of each exposure were flux calibrated, corrected to barycentric velocity, and corrected for spatial distortion (relative astrometric calibration). To achieve an absolute astrometric correction, we measured the centroid of one star in each exposure (using IRAF imexamine\footnote{IRAF was developed by the National Optical Astronomy Observatory \citep{1993ASPC...52..173T} and has been maintained by the IRAF community since 2017 (https://iraf-community.github.io/).}) and computed offsets relative to the ICRS coordinates of the Gaia DR2 tables \citep{GaiaDR2_astrometry}. The offsets and all processed pixel tables were then used to reconstruct a final cube for the field around Mk\,34, sampled at the standard MUSE NFM voxel size of $25.19$\,mas$\,\times\,25.42$\,mas$\,\times\,1.25$\,\AA. We also reconstructed images of the field in different broad-band filters.

Additionally, we  used  in this work data in NGC\,2070 obtained as part of the MUSE science verification and commissioning  of the wide-field-mode (MUSE-WFM) configuration in 2014 \citep{2014Msngr.157...13B}. MUSE-WFM provides a large field of view of 1'$\times$1', and a pixel scale of 0.2", but with a lower spatial resolution. The mosaic of NGC\,2070 obtained in 2014 \citep{2021A&A...648A..65C}  reached an image quality of 1". For details on the observations and our reduction of NGC\.2070 MUSE-WFM data, we refer to \cite{2018AA...614A.147C,2021A&A...648A..65C}.

\section{Searching for outflow fingerprints in the ISM}
\label{ISM}

The ISM surrounding Mk\,34 is inevitably influenced by the elevated ionizing spectrum of the massive binary system Mk\,34. To investigate this impact, we mapped the [N\textsc{ii}]\,$\lambda$\,6583 and [S\textsc{ii}]\,$\lambda$\,6717 emission lines around Mk\,34, searching for evidence of the influence of ionization on the ISM. We modeled both emission lines using a Gaussian function combined with a second-order polynomial. The polynomial component accounts for the continuum and the broad H$\alpha$ stellar wind component near Mk\,34. To enhance the signal-to-noise ratio, we averaged the flux within a $\pm$2 spatial pixel (spaxel) range during the field mapping. The systematic velocity of NGC\,2070 of 265\,km\,s$^{-1}$ \citep{2018AA...614A.147C} was taken into account before modeling the lines.

\begin{figure*}
        \centering
        \includegraphics[angle=0,width=0.49\textwidth]{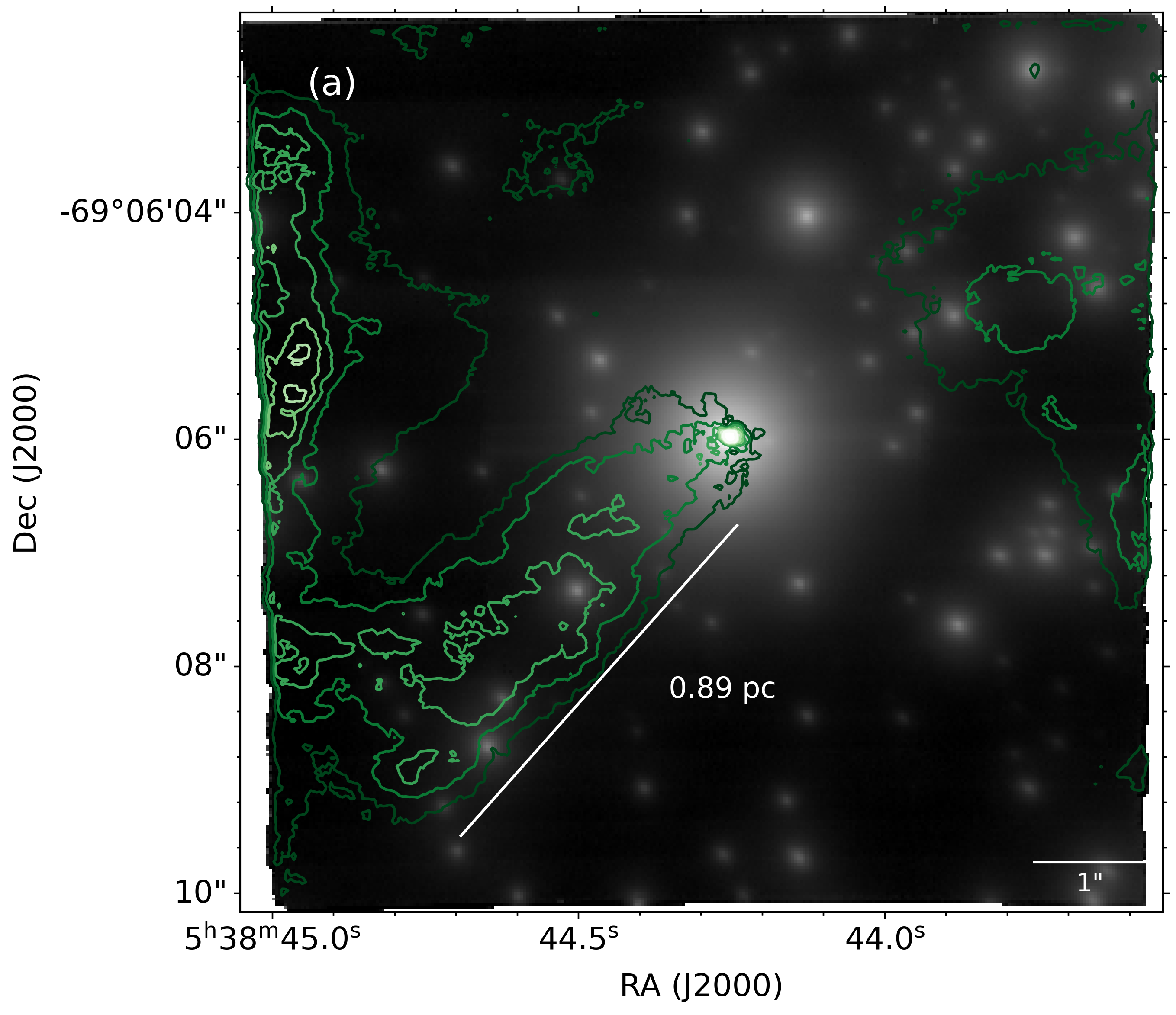}
        \includegraphics[angle=0,width=0.49\textwidth]{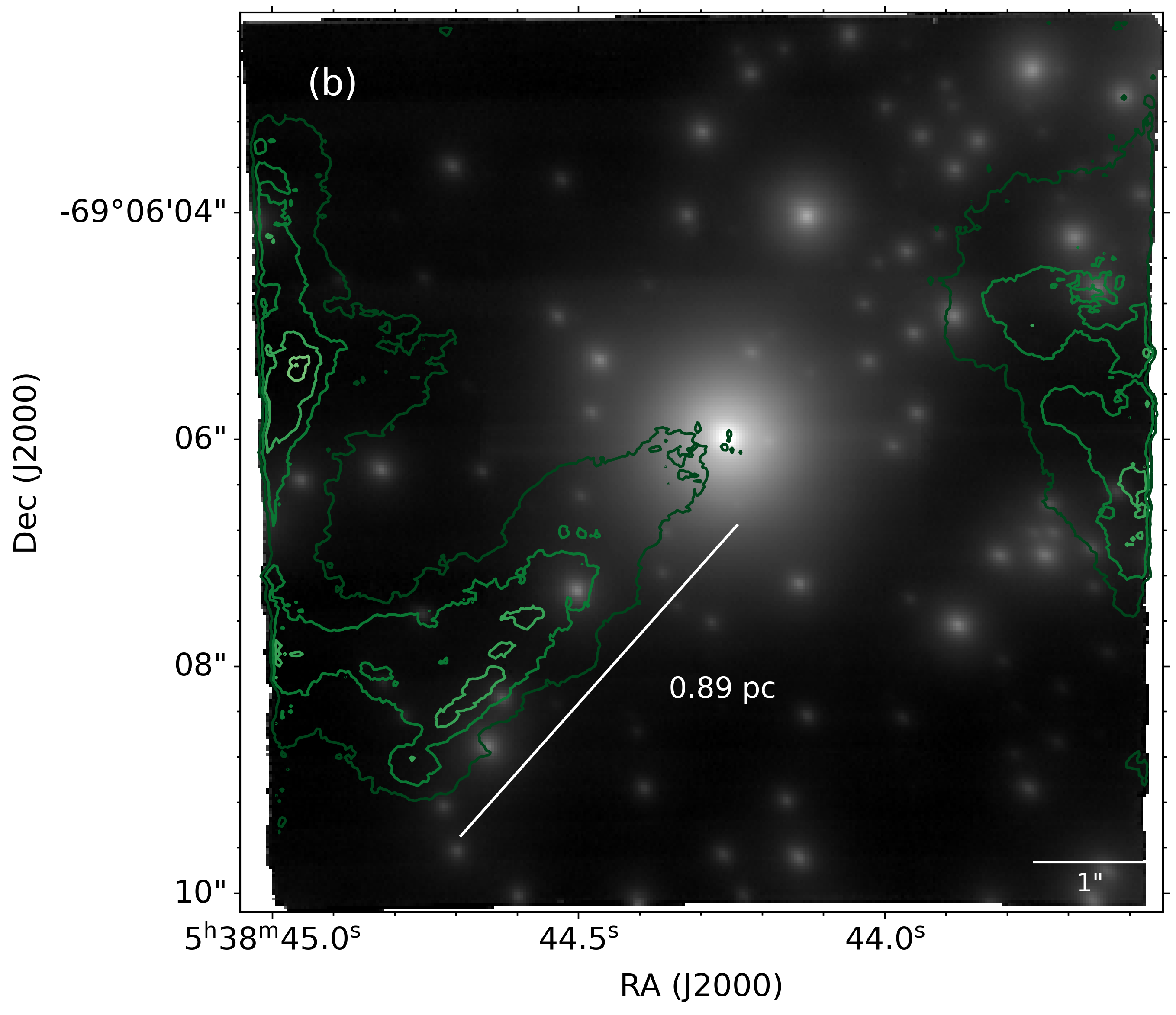}
        \caption{Emission line maps within the field of view of MUSE-NFM (approximately 7.5"$\times$7.5") around Mk\,34. Panel  \textit{a}: [N\textsc{ii}]\,$\lambda$\,6583 emission line map.  Panel  \textit{b}: [S\textsc{ii}]\,$\lambda$\,6717 emission line map.  We use the same intensity contours in both maps: contours range from $6\times10^{-19}$ to $2\times10^{-18}$ in steps of $2\times10^{-19}\,$erg$\,$s$^{-1}\,$cm$^{-2}\,$ per spaxel. The MUSE-NFM image in the Johnson V band is shown in the background of both panels.}
        \label{fig:blobs}
\end{figure*}

The emission from both ions shows a similar spatial distribution (Fig.~\ref{fig:blobs}), with filamentary structures that, in projection, appear to originate from Mk\,34 and extend towards the southeastern part of the field. These features suggest the presence of a plausible outflow originating from Mk\,34. Due to the consistent spatial distribution of both lines, and to avoid redundant information, the [N\textsc{ii}]\,$\lambda$\,6583 map is primarily used as the reference for discussion throughout this work.

Additionally, we explore the electron density  and the interstellar extinction maps around Mk\,34. The electron density was built based on  the  [S\textsc{ii}]\,$\lambda\lambda$6717,6731 ratio following the solution given for a three-level atom by \cite{1984MNRAS.208..253M} and assuming an electron temperature of T=10$^{4}$\,K (see also \citealt{2018AA...614A.147C}). The interstellar extinction c(H$\beta$) was obtained  from the H$\alpha$/H$\beta$ ratio, assuming Case B recombination theory for n$_{e}$\,=\,100\,cm$^{-3}$ and T=10$^{4}$\,K, that is, an intrinsic ratio of I(H$\alpha$)/I(H$\beta$)\,=\,2.86 \citep{1987MNRAS.224..801H}. We adopted a standard extinction law \citep{1989ApJ...345..245C} with RV\,=\,3.1 to ensure a direct comparison with previous results in NGC\,2070 using MUSE-WFM \citep{2018AA...614A.147C}.

\begin{figure*}
        \centering
        \includegraphics[angle=0,width=0.49\textwidth]{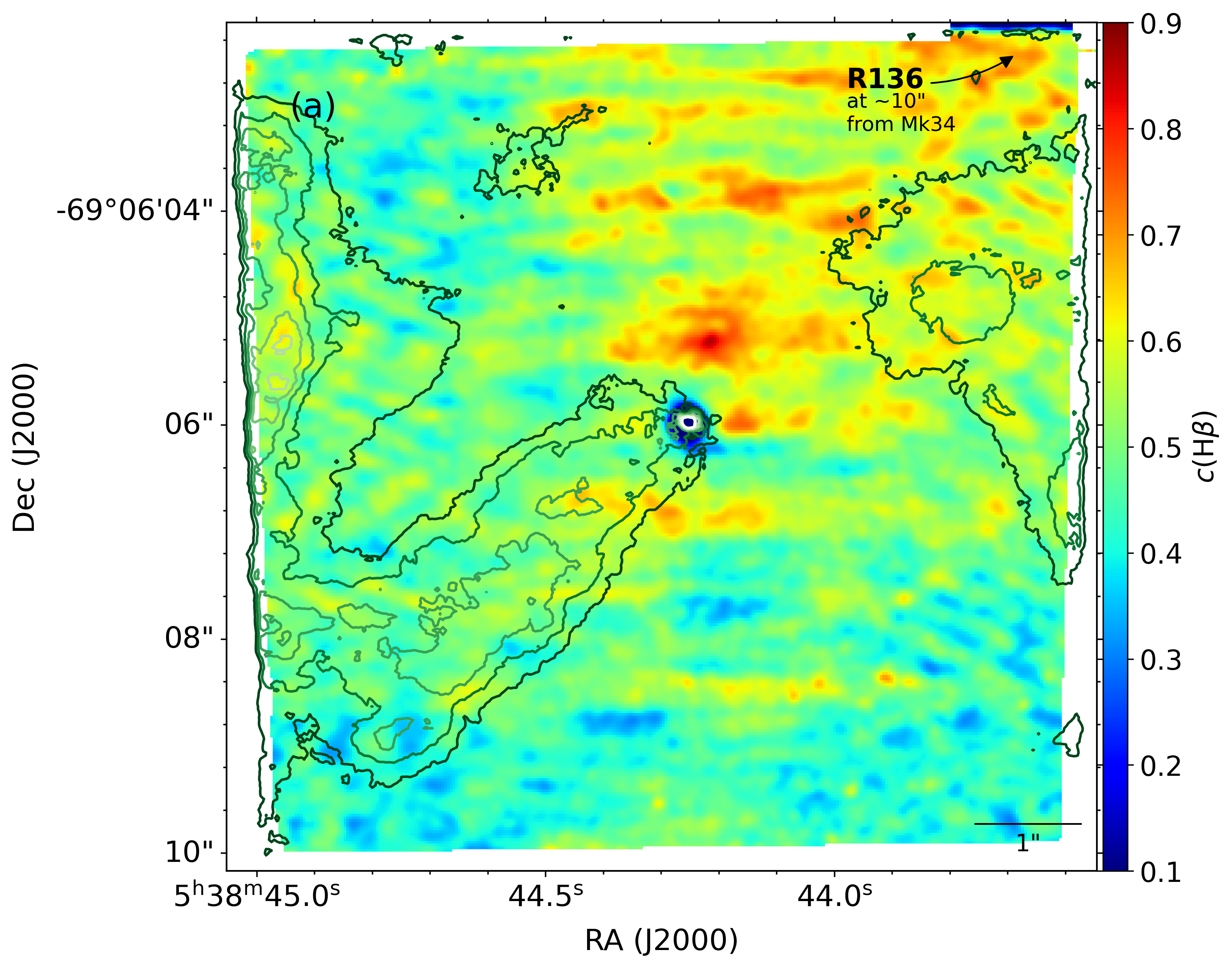}
        \includegraphics[angle=0,width=0.49\textwidth]{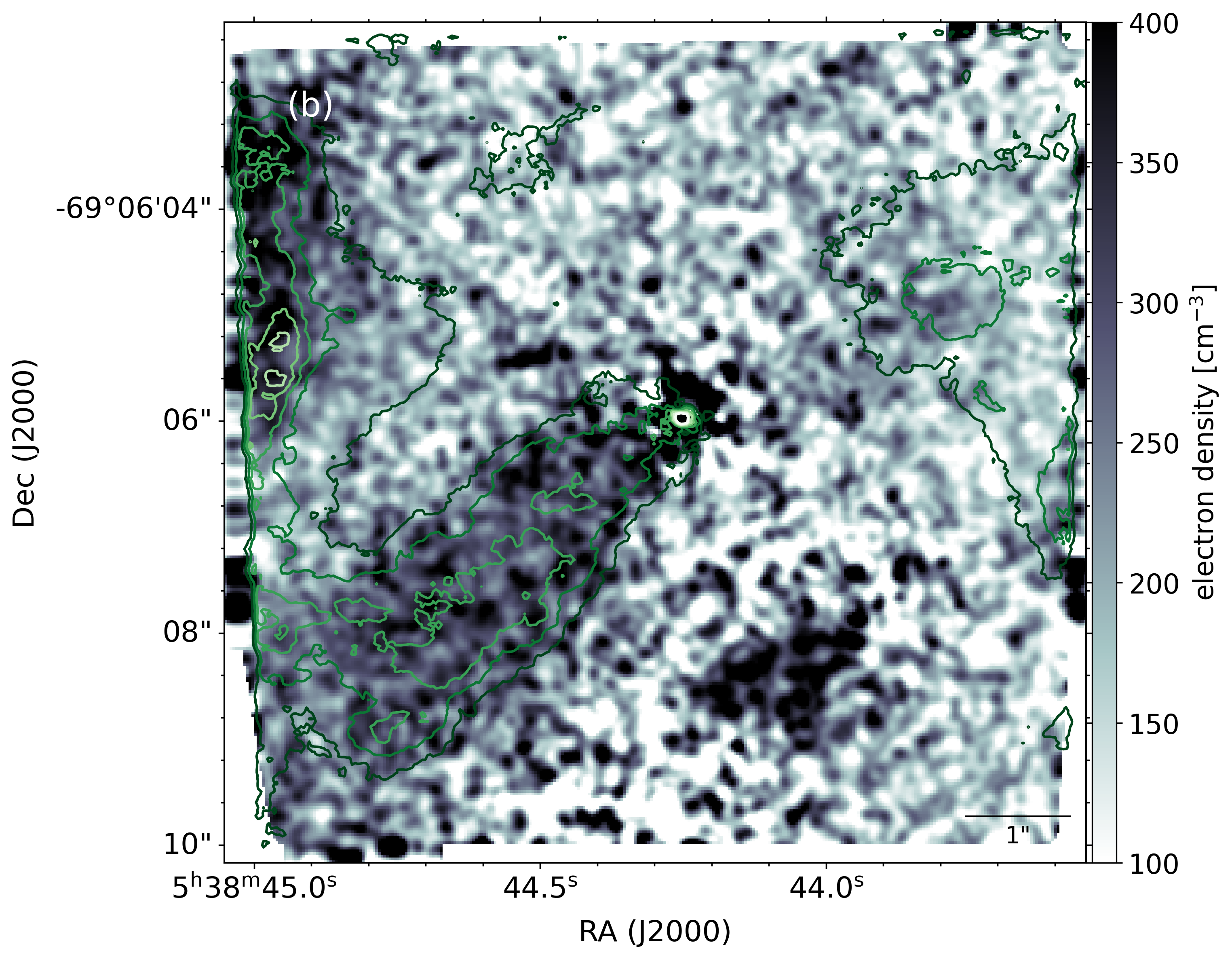}
        \caption{Reddening coefficient c(H$\beta$) (left panel) and electron density (right panel) maps. Both panels display the emission distribution 
        of [N\textsc{ii}]$\,\lambda 6583$ (green contours) described in Fig.~\ref{fig:blobs} as a reference and, 
        marking the proposed direction of the main outflow in Mk\,34. The distance to the center of  NGC\,2070, the cluster R136, is marked in panel \textit{a}. }
        \label{fig:ext}
\end{figure*}

The ISM electron density strongly marks the same [N\textsc{ii}] and [S\textsc{ii}] emission patterns, a cone structure apparently originating in Mk\,34 (panel \textit{b} in Fig.~\ref{fig:ext}). We do not find a symmetric counterpart of this cone structure on the northwest side of the field. However, the extinction map --retrieved using the ISM H$\alpha$ and H$\beta$ emission lines (panel \textit{a} in Fig.~\ref{fig:ext})--- reveals higher values in the northwest toward the center of the cluster. The larger extinction could partially obscure any  possible counterpart of the [N\textsc{ii}] contribution. Additionally,  the presence of the massive cluster R136 can also make it harder for any outflow to propagate toward it. A projected outflow would find an easier escape route moving away from R136, similar to our [N\textsc{ii}] detection. An analog to this scenario at a larger scale may be the outflows detected in the candidate Lyman continuum emitter Mrk\,71/NGC\,2366 \citep{2019A&A...623A.145M}.

\begin{figure}
        \centering
        \includegraphics[angle=0,width=0.5\textwidth]{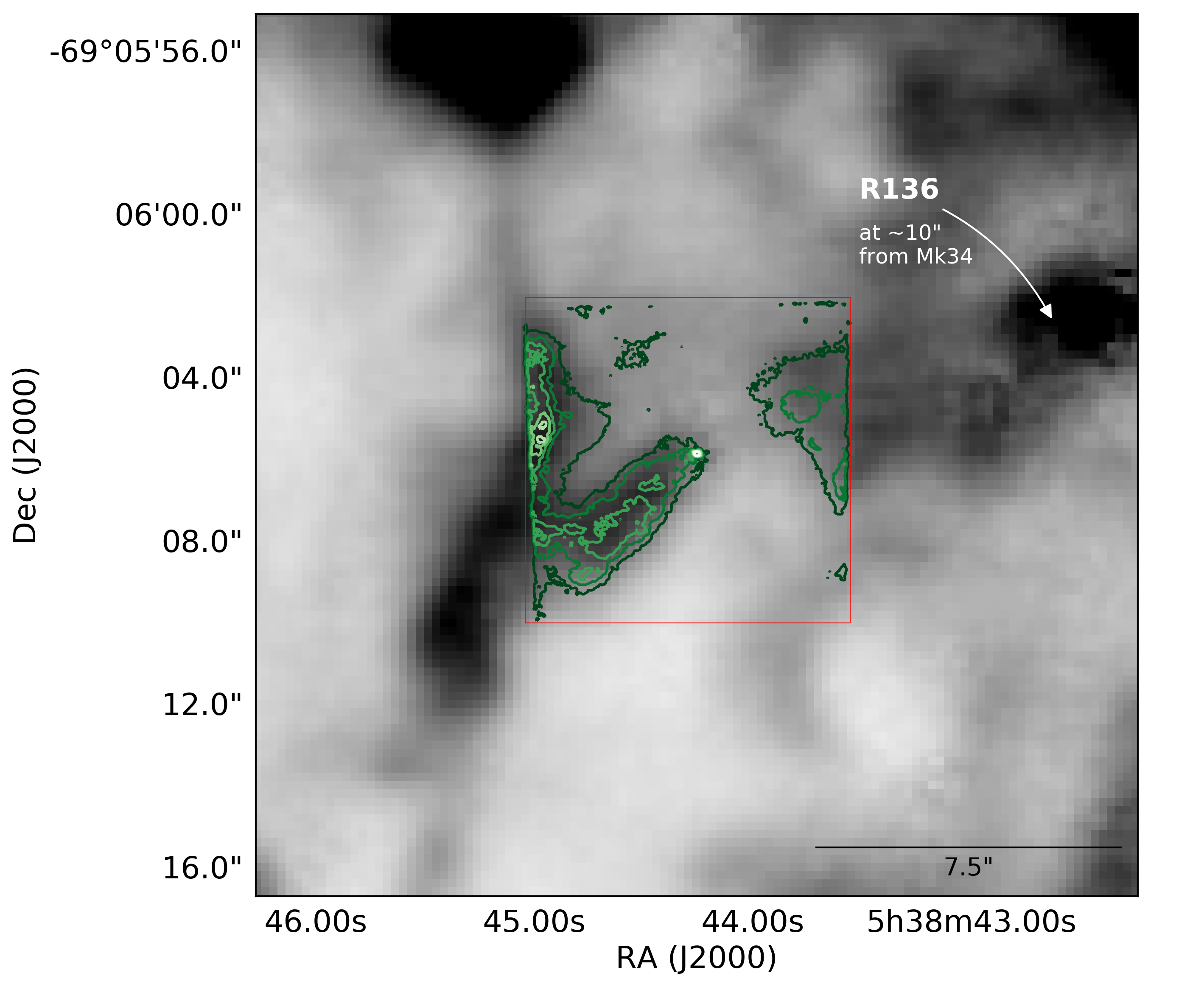}

        \caption{[N\textsc{ii}]$\,\lambda$\,6583 emission mapped by MUSE-WFM \citep{2018AA...614A.147C} around Mk\,34 with a spatial resolution of approximately 1". The [N\textsc{ii}]$\,\lambda$\,6583 emission mapped by MUSE-NFM is overplotted in green contours (see Fig.~\ref{fig:blobs}). The MUSE-NFM data sample the emission with a spatial resolution of approximately 80\,mas and a field of view of 7.5"$\times$7.5" (red square). The cluster R136 is indicated toward the northwest side of the field.}
        \label{fig:wfm}
\end{figure}

The [N\textsc{ii}] intensity map shows a narrow flow of material likely ejected from Mk\,34 at a projected distance of 0.89\,pc from the binary. Nevertheless, the emission close to the east edge of the MUSE-NFM field of view also suggests that the pattern is more complex. [N\textsc{ii}] intensity maps obtained with previous MUSE works \citep{2018AA...614A.147C} did not reach the spatial resolution of MUSE-NFM, but provide a larger view of the nitrogen emission map. Figure~\ref{fig:wfm} combines both observing modes of MUSE, WFM, and NFM, and shows a more complex scenario. We cannot discard the composition of different clouds in the final projected image. Nonetheless, the focus of the [N\textsc{ii}] emission around Mk\,34 is remarkable and points to the binary system Mk\,34 as the plausible source.

\begin{figure}
        \centering
        \includegraphics[angle=0,width=0.5\textwidth]{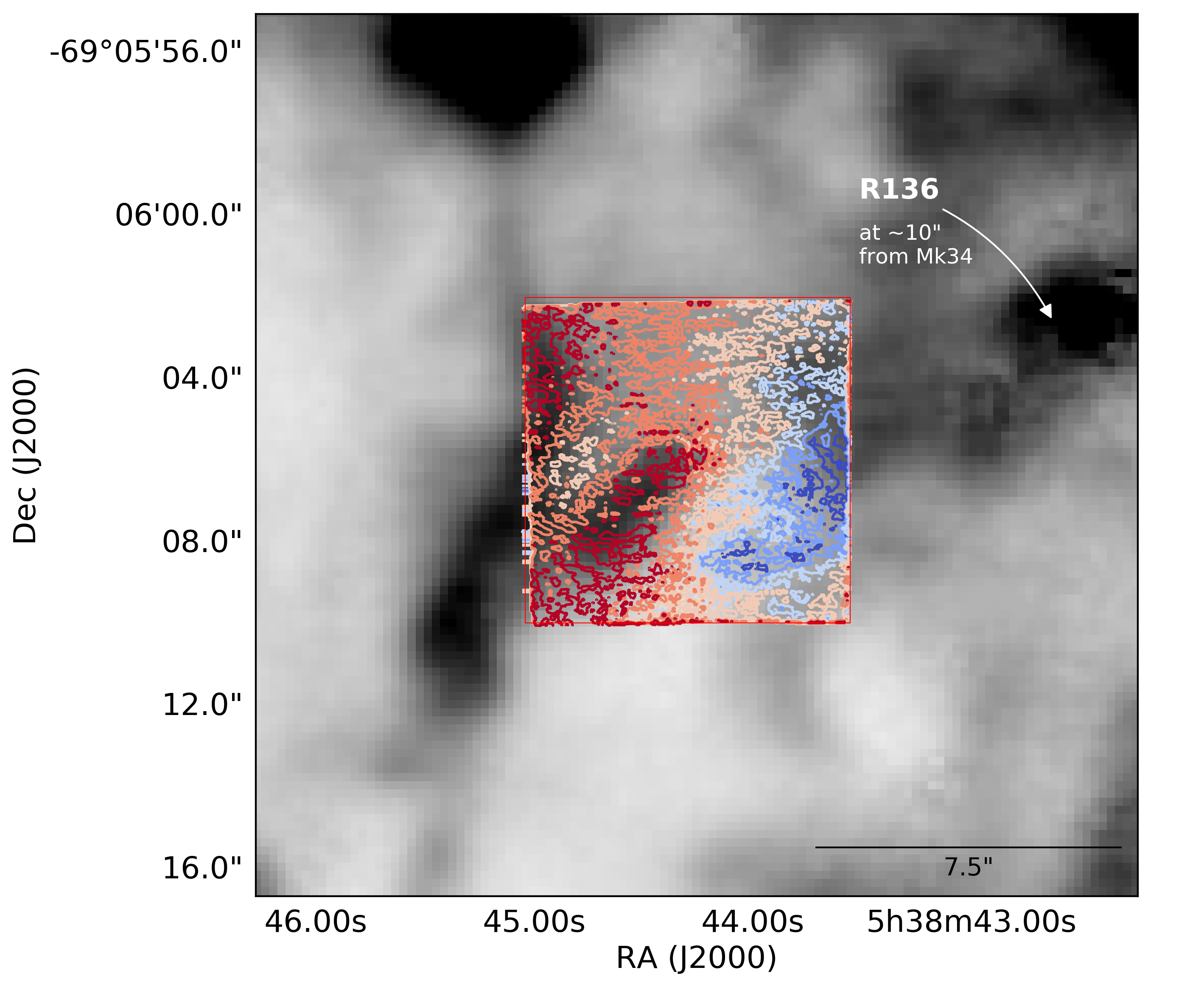}

        \caption{[N\textsc{ii}]$\,\lambda$\,6583 emission mapped as described in Fig.~\ref{fig:wfm}. The [N\textsc{ii}] radial velocity mapped  by MUSE-NFM is overplotted in red and blue contours. Contours range from -30\,km\,s$^{-1}$ (blue) to 30\,km\,s$^{-1}$ (red) in steps of 10\,km\,s$^{-1}$ (NGC\,2070 systemic velocity has been subtracted). See Fig.~\ref{fig:wfm} as a reference for the additional labels. }
        \label{fig:wfm_velo}
\end{figure}

Figure~\ref{fig:wfm_velo} displays the radial velocity map of [N\textsc{ii}] overlaid on the overall view provided by the MUSE-WFM, as shown in Fig.~\ref{fig:wfm}. The filament, traced by  [N\textsc{ii}], is redshifted with respect to the systemic velocity of NGC\,2070, moving away from us in the southeast part of Mk\,34. We measure a projected velocity of approximately 30\,km\,s$^{-1}$ in the peak of the [N\textsc{ii}] velocity map. A counterpart of blueshifted material is spotted on the west side of the image reaching velocities of a similar order (-40\,km\,s$^{-1}$).

\section{Revisiting Mk\,34 and exploring variability}
\label{epoch}

\begin{figure*}
        \centering
        \includegraphics[angle=0,width=0.49\textwidth]{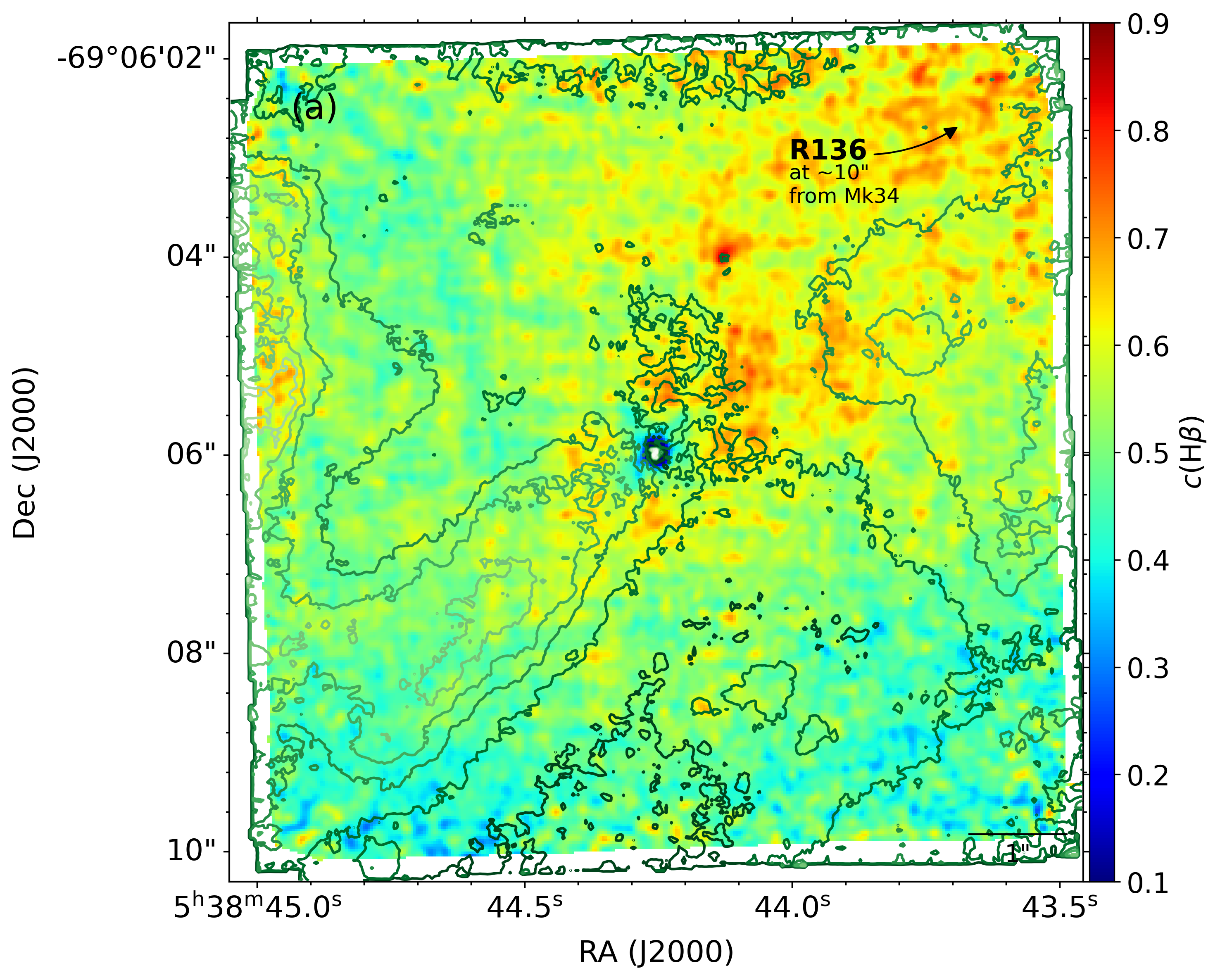}
        \includegraphics[angle=0,width=0.49\textwidth]{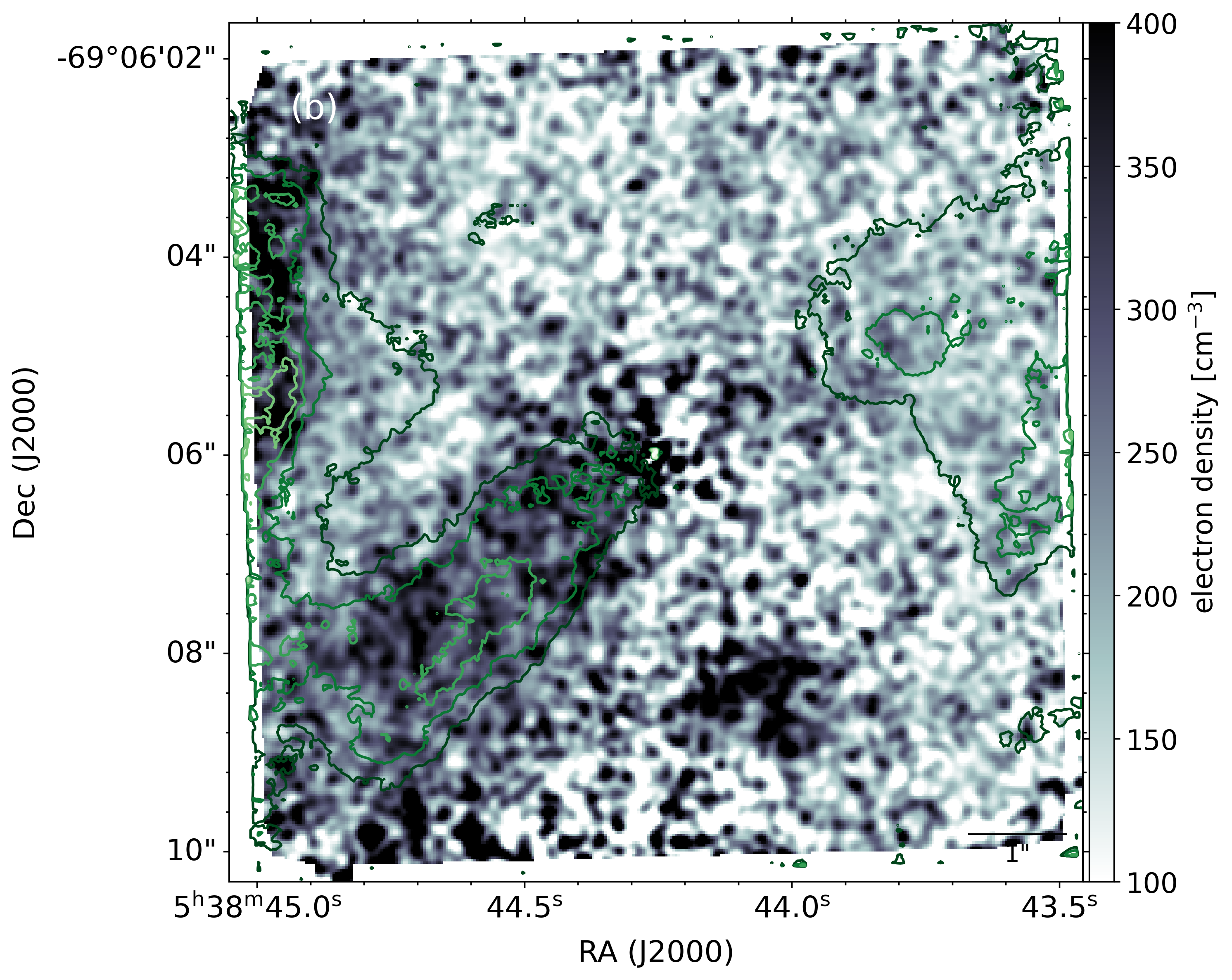}
        \label{fig:2021}
        \caption{Reddening coefficient c(H$\beta$) (left panel) and  electron density (right panel) maps extracted from the 2021 epoch. In the left panel, the fainter contours of [N\textsc{ii}]\,$\lambda $\,6583 emission are delineated (from $2\times10^{-19}$ to $2\times10^{-18}\,$erg$\,$s$^{-1}\,$cm$^{-2}\,$ in increments of $2\times10^{-19}\,$erg$\,$s$^{-1}\,$cm$^{-2}\,$ per spaxel) in comparison with the depiction in Fig.~\ref{fig:blobs}. This adjustment was made to accentuate the observed correlation between large c(H$\beta$) values and the absence of [N\textsc{ii}] emission in the northwest of Mk\,34.        }
\end{figure*}

Mk\,34 was observed again in December 2021, with a time interval of 745 days between the two epochs. The [N\textsc{ii}]\,$\lambda$\,6583 and [S\textsc{ii}]\,$\lambda$\,6717 intensity maps in  2021 display the same outflow pattern as in 2019. The electron density and extinction maps also support the same previous findings, a cone-like structure apparently emerging from Mk\,34 in the southeast and more significant extinction in the northwest part toward the center of NGC\,2070 (see Fig.~\ref{fig:2021}). Fainter [N\textsc{ii}] contours are drawn in the extinction map in Fig.~\ref{fig:2021}, highlighting the correlation between higher extinction and the lack of an [N\textsc{ii}] emission counterpart in the northwest side of Mk\,34.

\begin{figure}
        \centering
        \includegraphics[angle=0,width=0.49\textwidth]{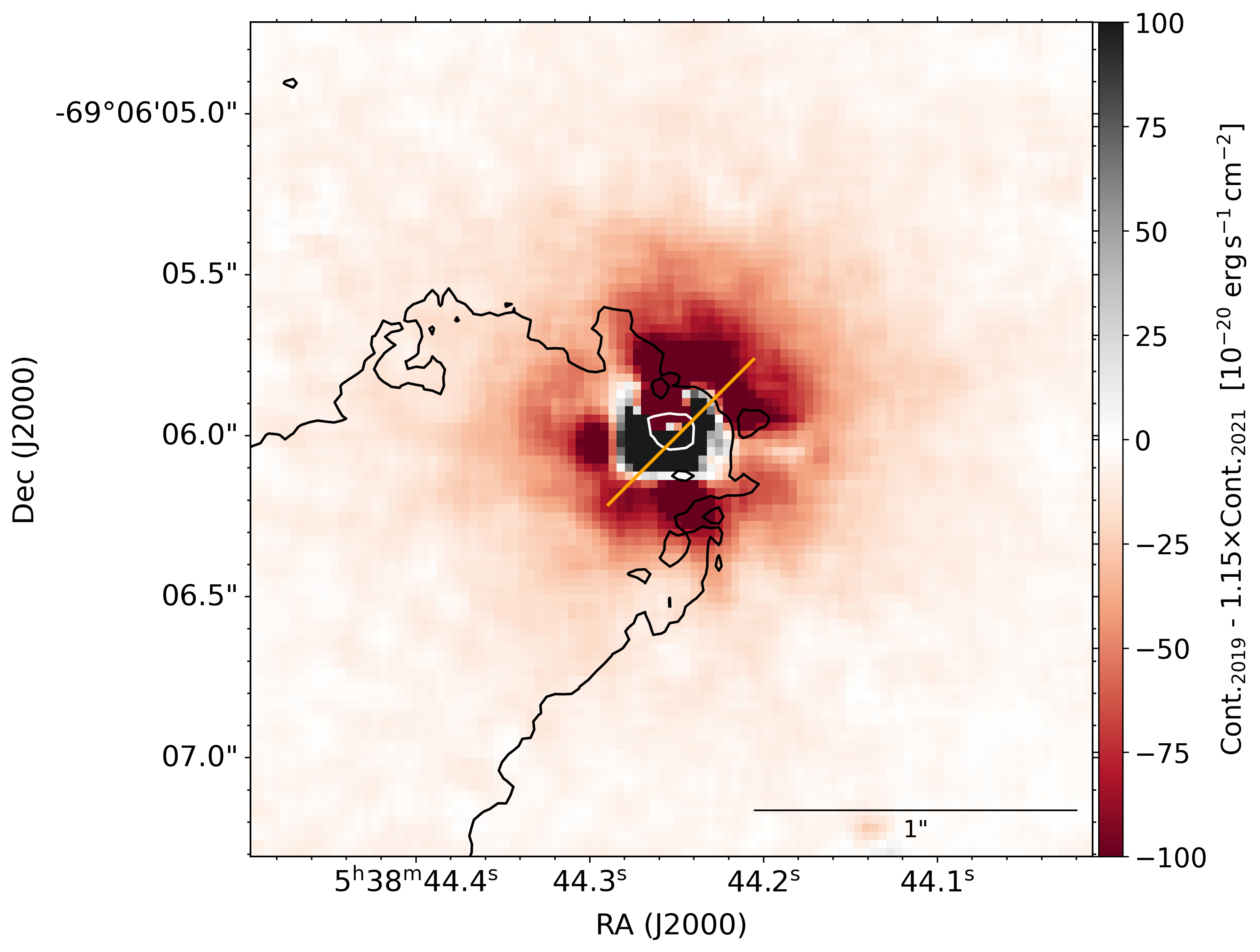}
        \caption{Difference in the continuum flux around [N\textsc{ii}]\,$\lambda$\,6583 between the observations in 2019 and 2021. A factor of 1.15 was applied to the 2021 observations to correct for differences in the flux calibrations (see Sect.~\ref{epoch}). The proposed flow direction is indicated by a solid orange line, along with the edges in the [N\textsc{ii}]\,$\lambda $\,6583 contours displayed in Figs.~\ref{fig:blobs} and \ref{fig:2021} to highlight the path within the ISM.}        
        \label{fig:cont}
\end{figure}

The difference between the two observing epochs shows a systematic flux discrepancy between all the sources in the field. We attribute this to an issue in the flux calibration. Before comparing the two epochs, it is necessary to put the fluxes and positions of the stars in a mutual reference system. We use the objects detected by the Gaia \citep{GaiaDR2_astrometry} survey around Mk\,34 as calibration anchors within the MUSE-NFM field of view, finding ten stars in total. First, we checked the positions in both data sets and found variations of less than 0.5 pixels on average. Based on this result, we conclude that a correction in the astrometry is not needed. Second, we measured the continuum flux around [N\textsc{ii}]\,$\lambda$\,6583  for the ten targets and find that we need a systematic correction of 1.15 in the flux to place the fluxes of the ten reference sources in the two epochs into the same reference system. Within several arcseconds around Mk\,34, we do not detect significant quantitative differences between the two epochs (Fig.~\ref{fig:cont}). A closer examination of Mk\,34 at a scale of less than 1", using the continuum flux around the [N\textsc{ii}]\,$\lambda$\,6583 emission line, reveals differences that may be associated with  short-term events occurring very close to the star. Moreover, the difference between the two images highlights a diagonal pattern, extending from southeast to northwest (marked in Fig.~\ref{fig:cont}), which aligns with the proposed [N\textsc{ii}] and [S\textsc{ii}] outflow maps.

\begin{figure*}
        \centering
        \includegraphics[angle=0,width=0.49\textwidth]{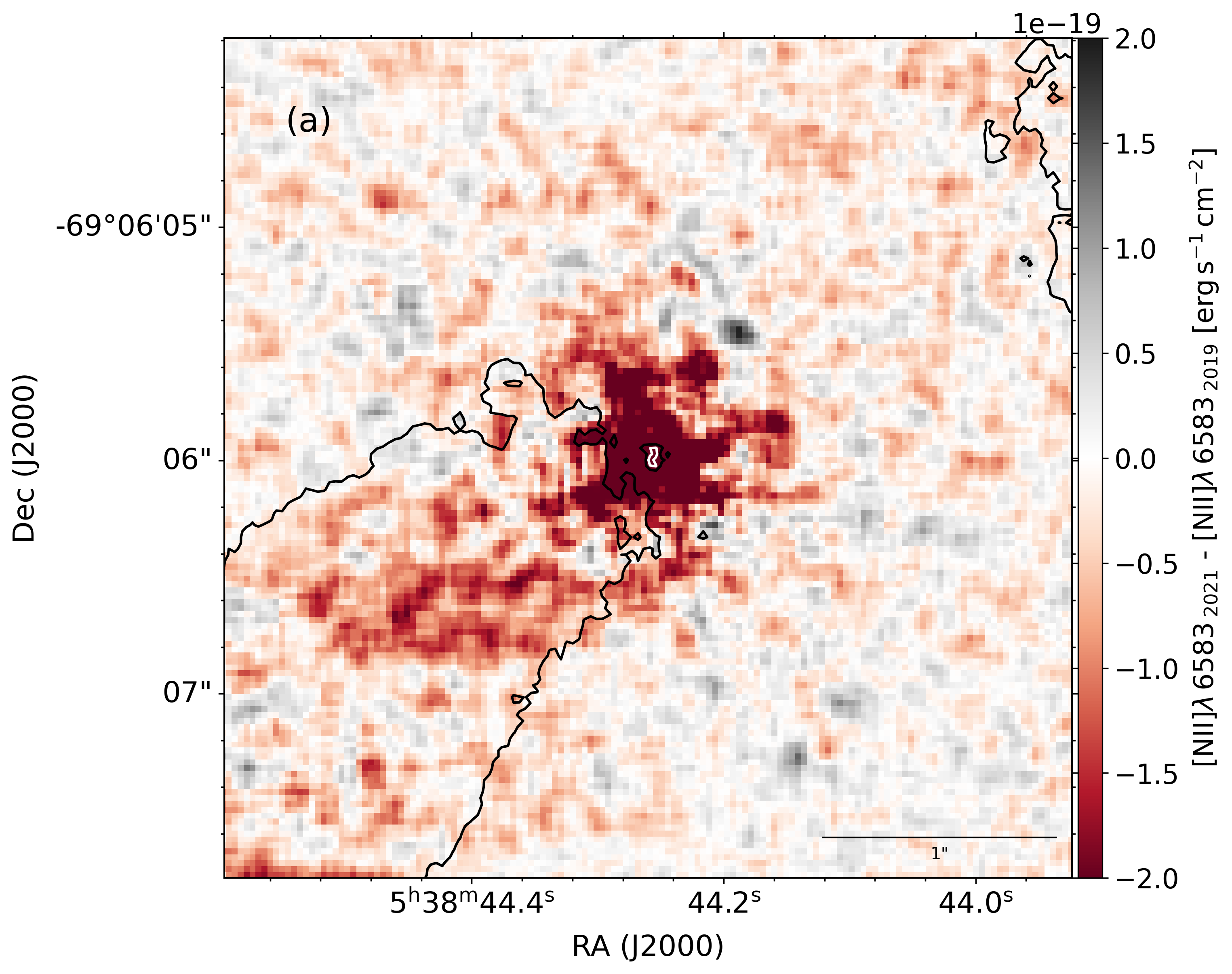}
        \includegraphics[angle=0,width=0.49\textwidth]{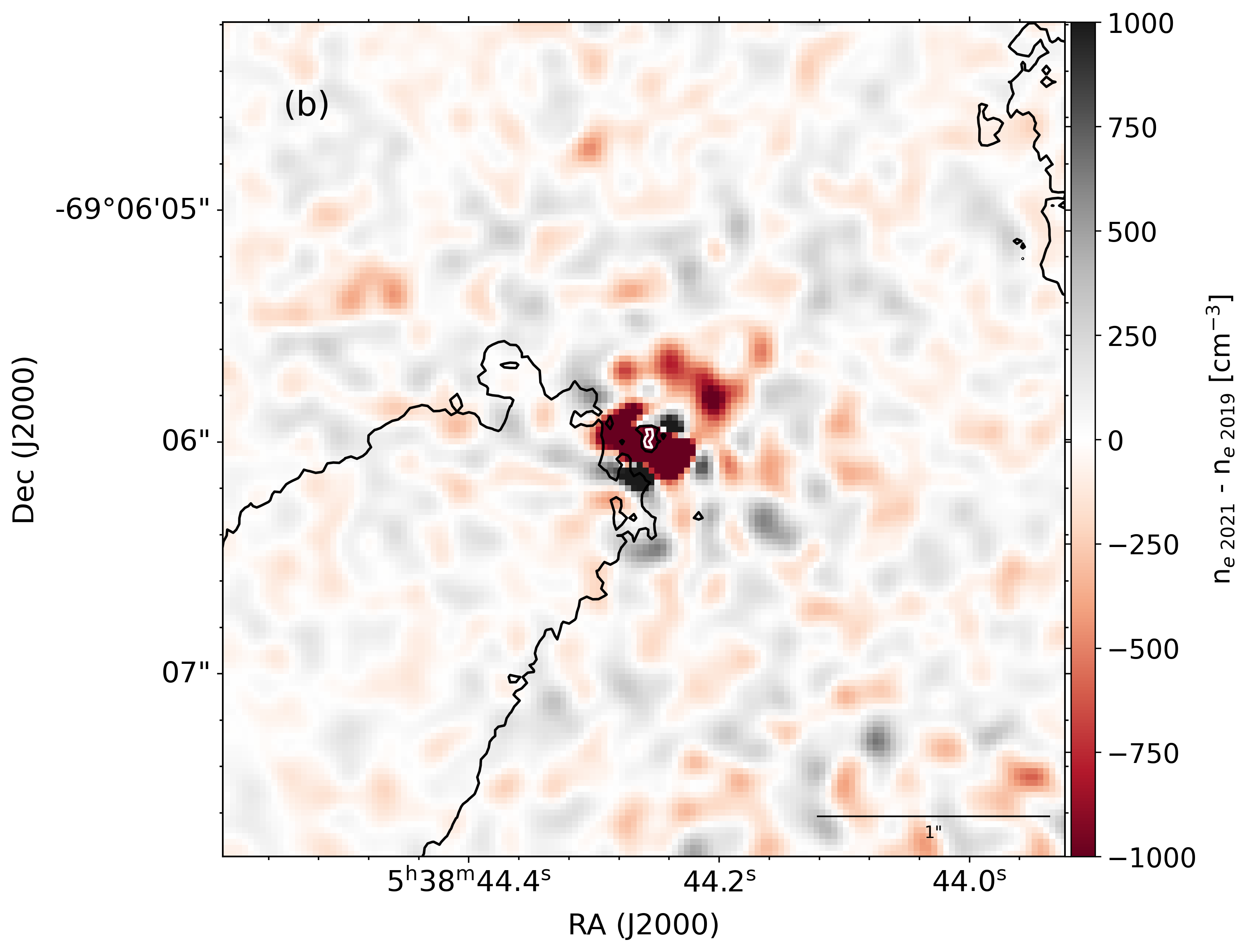}
        \includegraphics[angle=0,width=0.49\textwidth]{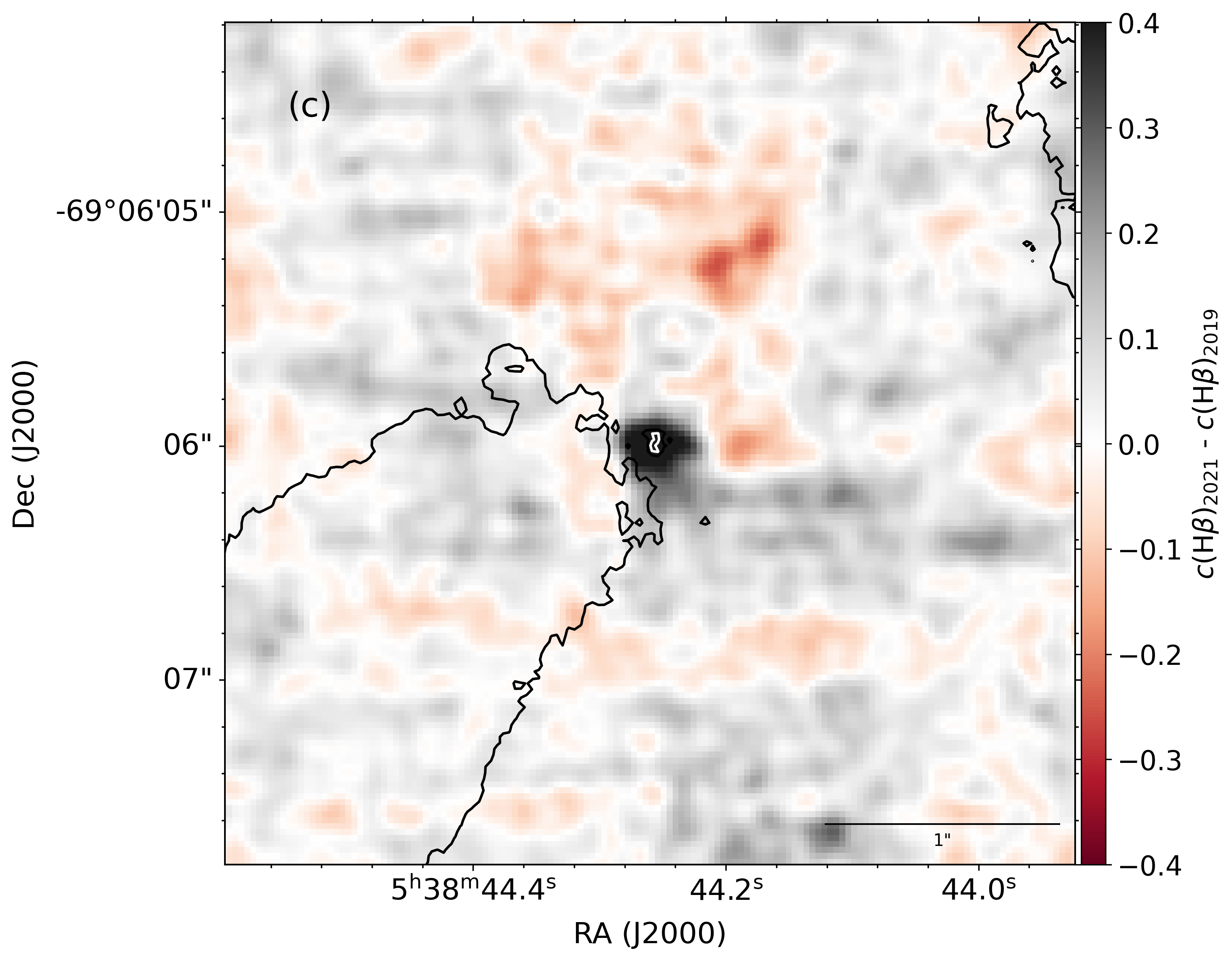}
        \includegraphics[angle=0,width=0.49\textwidth]{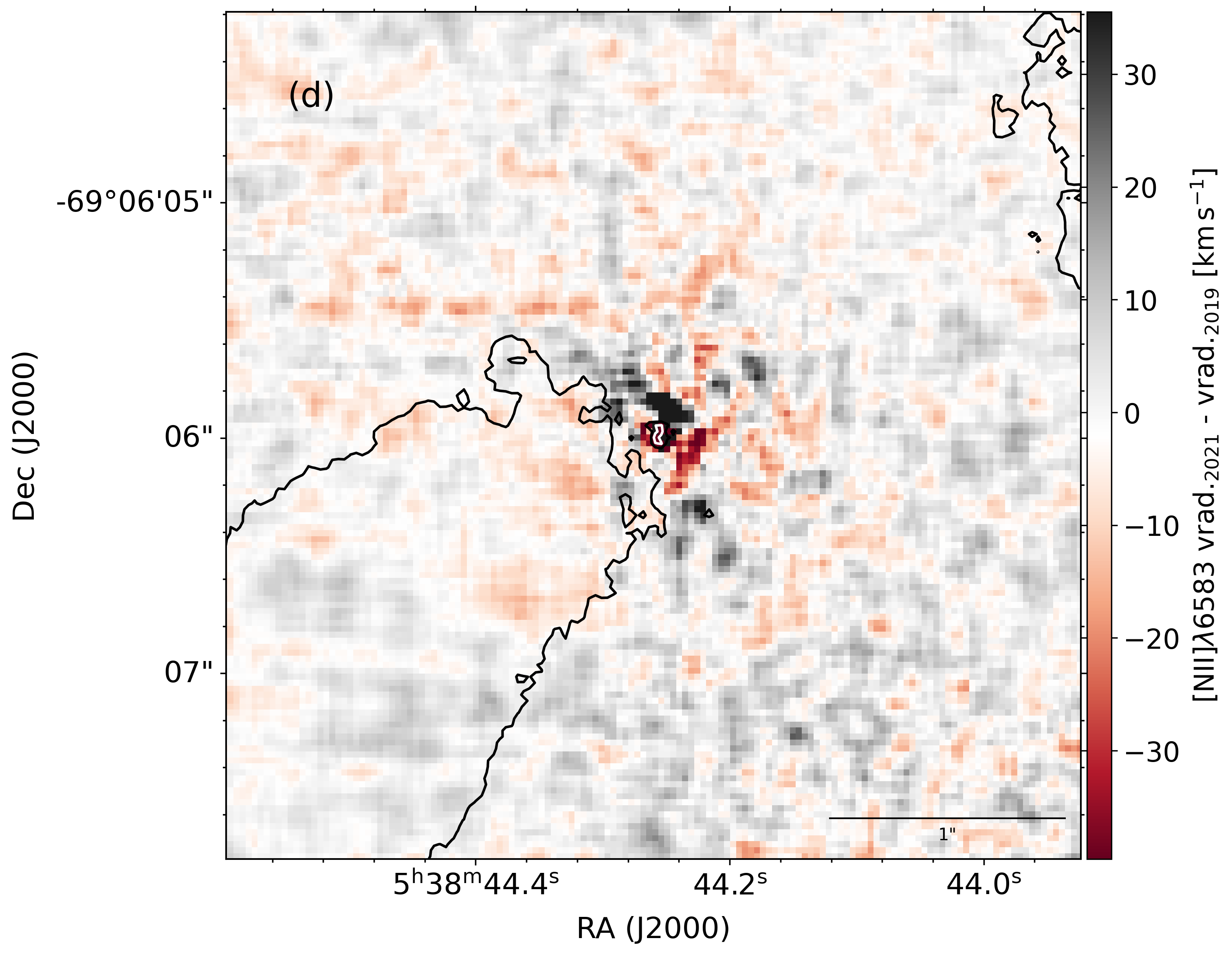}
        \caption{Comparison of emission and physical properties between the 2019 and 2021 epochs.
        Panel \textit{a}: Differences in [N\textsc{ii}]\,$\lambda$\,6583 between the two epochs. Panel \textit{b}: Difference in electron density measured in 2019 and 2021. Panel \textit{c}: Difference in extinction measured in 2019 and 2021. Panel \textit{d}: Difference in [N\textsc{ii}]\,$\lambda$\,6583 radial velocities measured in 2019 and 2021. The path of [N\textsc{ii}]\,$\lambda$\,6583 intensity is highlighted by the edges in the [N\textsc{ii}]\,$\lambda $\,6583 contours displayed in Figs.~\ref{fig:blobs} and \ref{fig:2021}, marking the detection within the ISM across all four panels. }
        \label{fig:2021_2019}
\end{figure*}

Figure~\ref{fig:2021_2019} illustrates the differences between the two epochs for the properties analyzed in this study. The intensity of the [N\textsc{ii}] $\lambda$\,6583 line shows stronger emission in 2019 around Mk\,34 and along the ISM outflow path compared to the 2021 data. The electron density and extinction exhibit significant changes primarily in the region closer to Mk\,34. We observed substantial differences in density at the location of  Mk\,34, with the electron density in 2019 being significantly higher than in 2021. However, we must approach these large values with caution, as the three-level atom solution provided by \cite{1984MNRAS.208..253M} may not be applicable in the environment surrounding a massive and hot binary system like Mk\,34. Additionally, assuming an electron temperature of T=10$^{4}$ K near Mk\,34 may be unrealistic. The final panel in Fig.~\ref{fig:2021_2019} does not reveal any clear pattern in the [N\textsc{ii}]\,$\lambda$\,6583 radial velocities between the two epochs, aside from a few small features near the binary system. These velocity measurements could be influenced by the significant H$\alpha$ contribution.

\section{Modeling and extraction of the point spread function}
\label{psf}

We analyzed the residuals of the point spread function (PSF) beneath the strong flux profile of Mk\,34 in the MUSE data to explore the intense activity of the binary system within 0.5" and to investigate potential the links to ISM emissions. Additionally, we examined how small-scale variations might be influenced by the effects of the AO-corrected PSF. The PSF of Mk\,34 was modeled and subtracted at each wavelength of the datacube individually for both epochs. The modeling was performed using the MAOPPY (Modelization of the Adaptive Optics PSF in Python) library\footnote{https://gitlab.lam.fr/lam-grd-public/maoppy}, which is specifically designed for modeling PSFs in adaptive optics(AO)-assisted observations. The MAOPPY models have been tested on MUSE data taken in AO narrow-field mode, achieving a relative error of less than 1\% for both simulated and experimental data \citep{2019A&A...628A..99F}.

\begin{figure*}
        \centering
        \includegraphics[angle=0,width=0.49\textwidth]{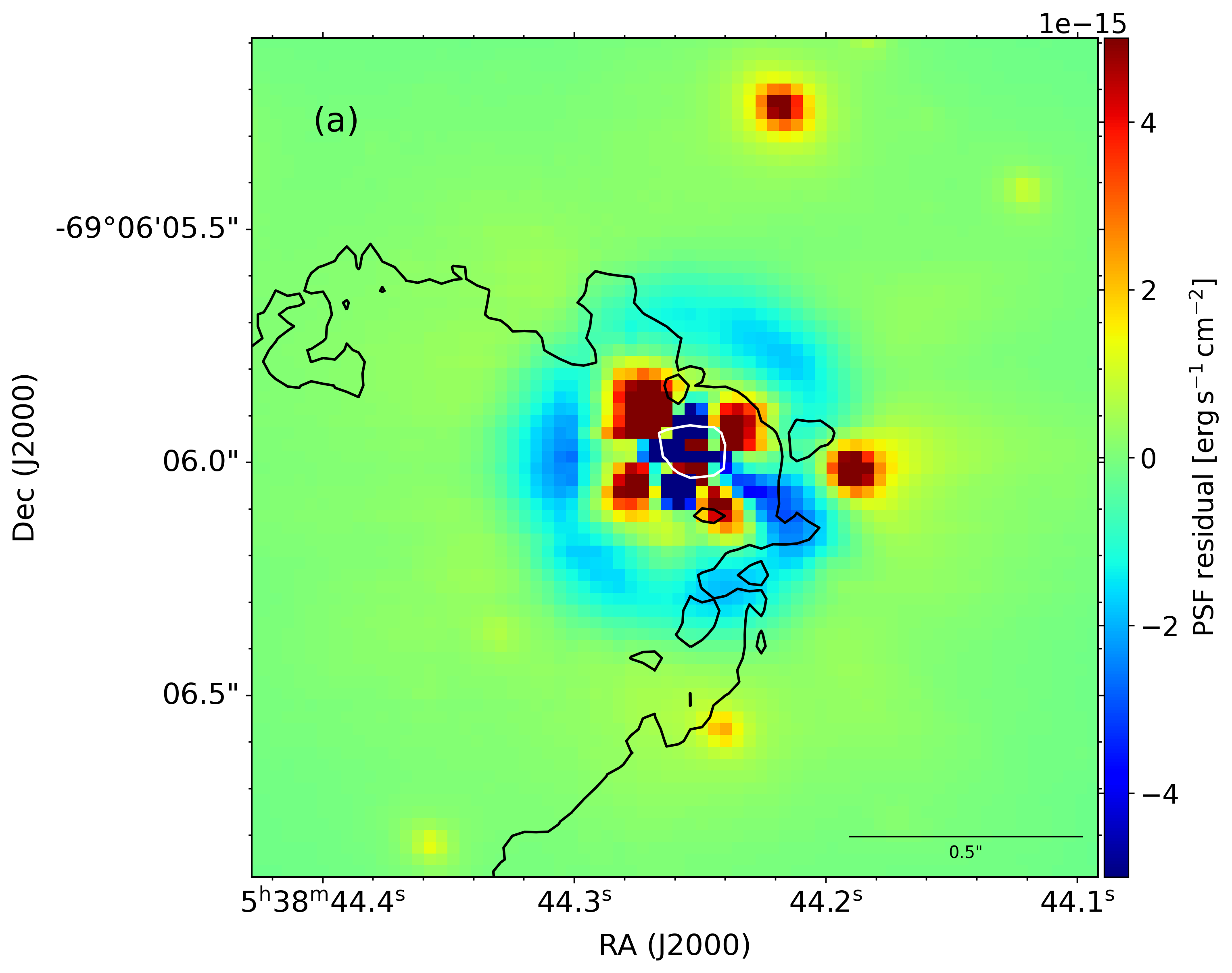}
        \includegraphics[angle=0,width=0.49\textwidth]{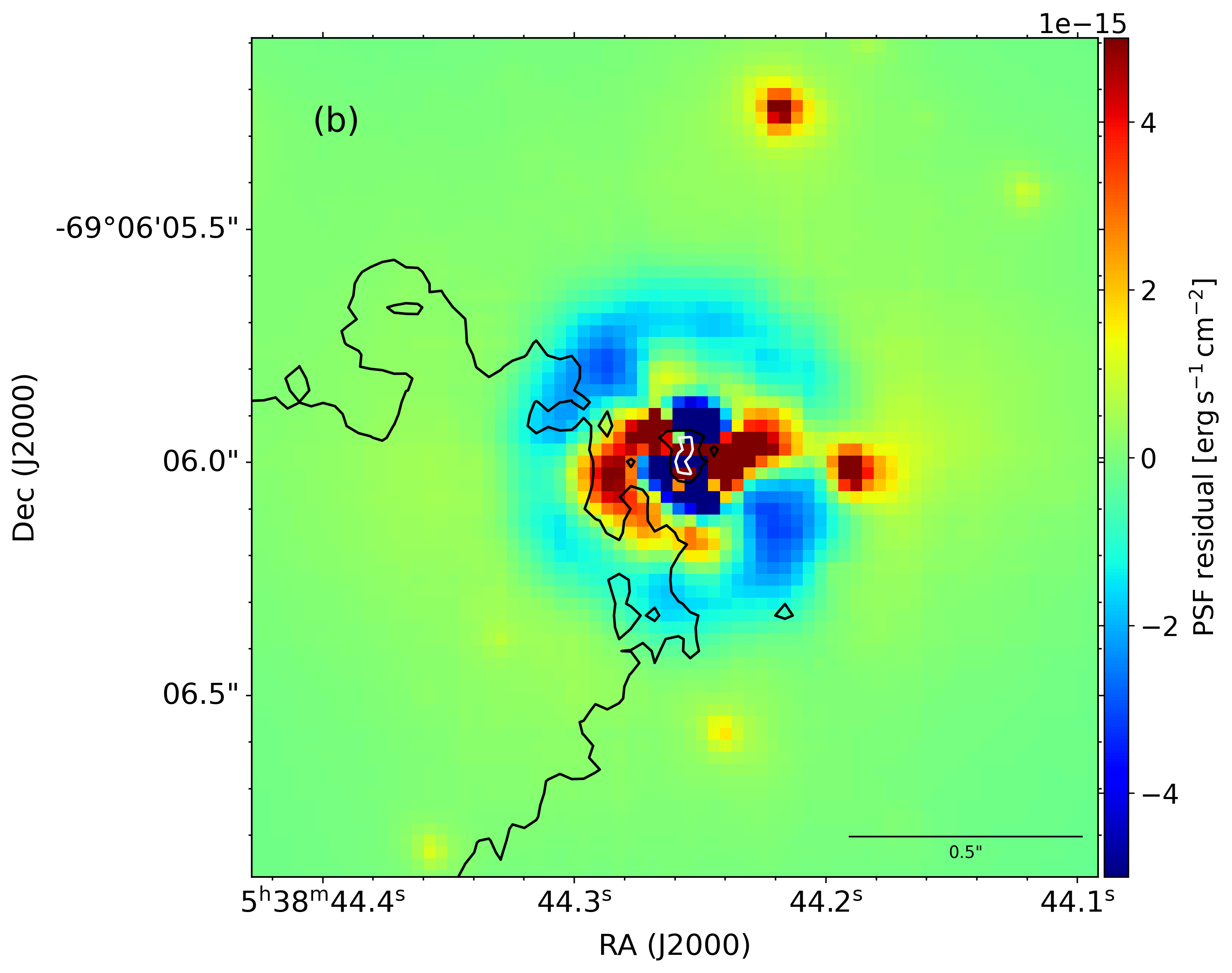}
        \includegraphics[angle=0,width=0.49\textwidth]{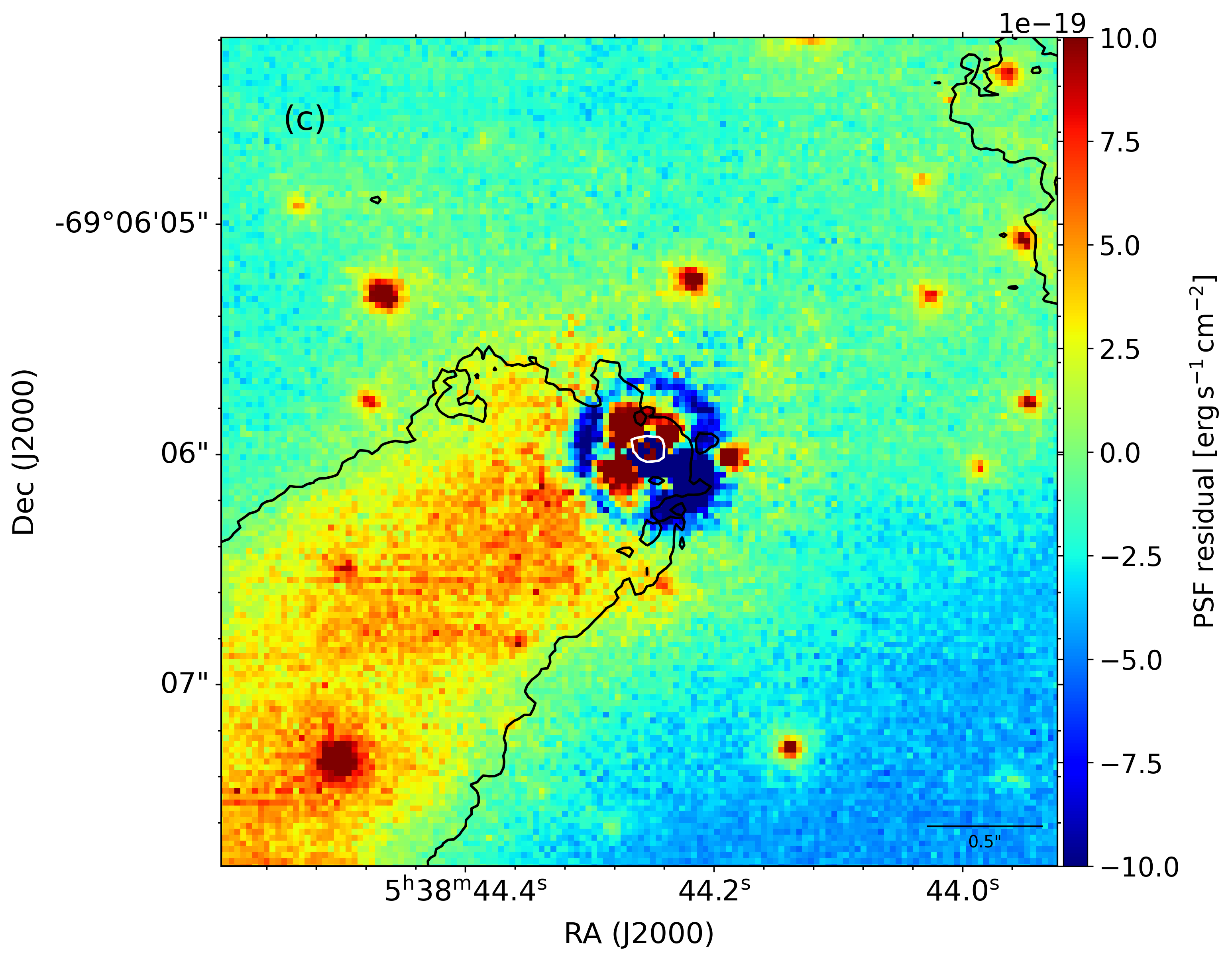}
        \includegraphics[angle=0,width=0.49\textwidth]{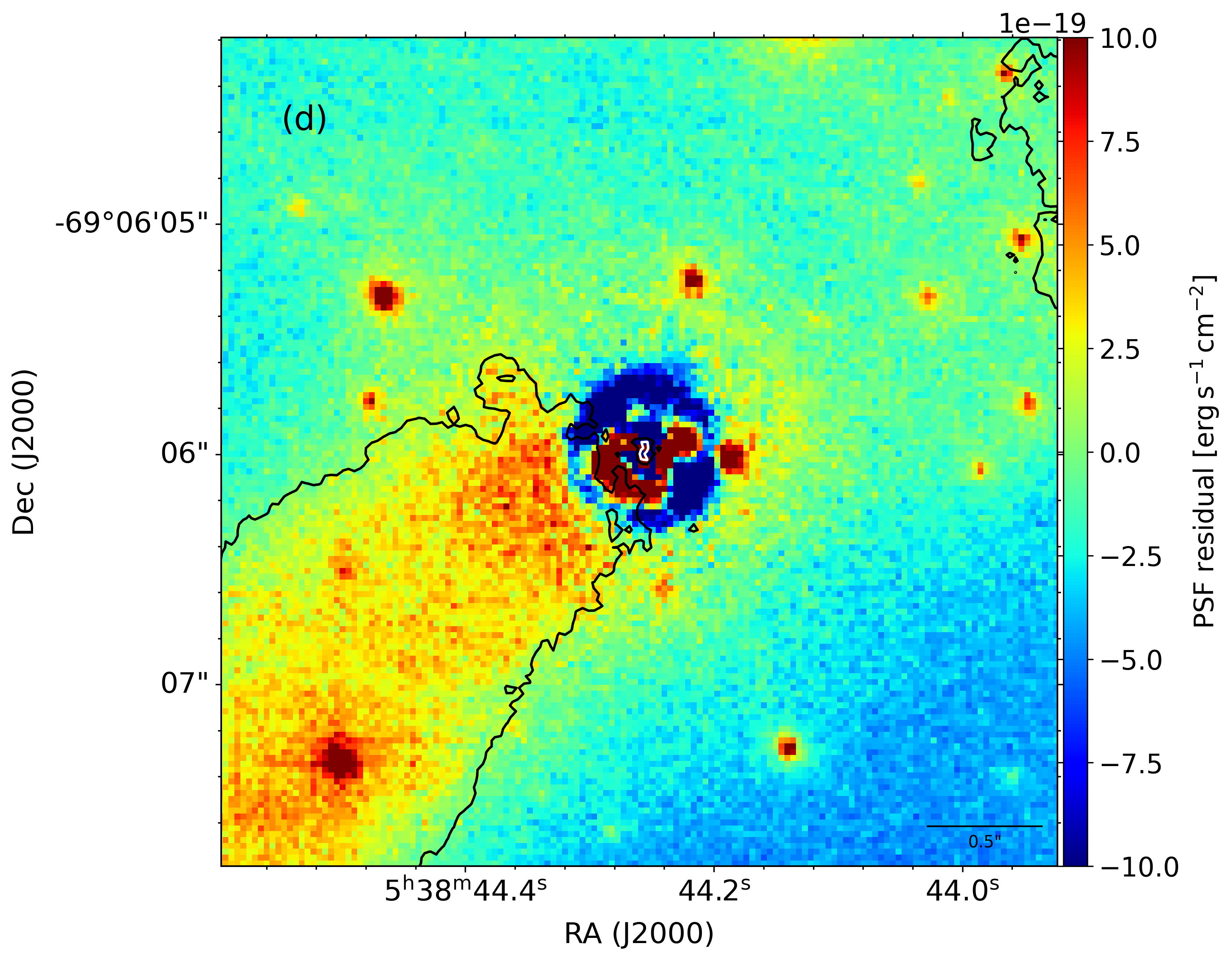}
        \caption{ PSF residual maps. Panel \textit{a}: Total PSF residual in 2019. Panel \textit{b}: Total PSF residual in 2021.  Panel \textit{c}: PSF residual in 2019 at the position of [N\textsc{ii}]\,$\lambda$\,6583. Panel \textit{d}: PSF residual in 2021 at the position of [N\textsc{ii}]\,$\lambda$\,6583. The path of [N\textsc{ii}]\,$\lambda$\,6583 intensity is highlighted by the edges in the [N\textsc{ii}]\,$\lambda $\,6583 contours displayed in Figs.~\ref{fig:blobs} and \ref{fig:2021}, marking the detection within the ISM across all four panels.} 
        \label{fig:PSFs}
\end{figure*}

The total PSF residual in 2019, obtained after adding all the layers in the datacube, shows four main contributions around the core of Mk\,34, where the PSF model fit by MAOPPY under-subtracts the data. In contrast, in the central part and the region around these four points, the model produces a negative residual (Fig.~\ref{fig:PSFs}).  This negative feature resembles the residual maps reported by \cite{2019A&A...628A..99F} as part of their characterization of the MUSE-NFM PSF. The PSF residual in 2021 displays a similar average pattern to that in 2019; however, the details in the core change. The four contributions are replaced by two positive regions, primarily along the east--west axis of Mk\,34. It is worth mentioning here again that the 2021 data were obtained with four rotation settings (see Sect.~\ref{data}), which could influence the differences observed between the two epochs.

Figure~\ref{fig:PSFs} also displays the residual at the wavelength of [N\textsc{ii}]\,$\lambda$\,6583 after accounting for the systemic velocity of NGC\,2070. The outflow described in Sect.~\ref{ISM} is visible in both epochs. Only the PSF of Mk\,34  was modeled and subtracted in the present work. The remaining fainter stars, which are still visible in Fig.~\ref{fig:PSFs}, were not included in this exercise.

We do not find a clear connection between the PSF residuals near Mk\,34 and the ISM outflow. Although the positive pattern observed in 2021 may align with the direction indicated by the [N\textsc{ii}] and [S\textsc{ii}] intensity maps, evidence from a single epoch is insufficient to draw a conclusion. The PSF was modeled for each wavelength individually, revealing very similar patterns across the entire MUSE datacube in the 2019 and 2021 datacubes. This consistency adds robustness to the maps shown in Fig.~\ref{fig:PSFs}. We may be observing events near Mk\,34 that vary over short periods and potentially relate to the 155\,day period reported by \cite{2018MNRAS.474.3228P}. We tentatively hypothesize that we are witnessing material expelled as a result of interactions between these two very massive stars. While this scenario could be further explored with observations taken at the appropriate cadence, it cannot be confirmed with the current data.

\section{Mapping the outflow regions using the BPT diagram}
\label{bpt}

The BPT diagram is commonly employed to distinguish sources based on the nature of the ionizing origins, such as star-forming galaxies and active galactic nuclei \citep{1981PASP...93....5B}, as well as to differentiate between shocks and photoionized gas (e.g., HII regions from planetary nebulae and supernova remnants; \citealt{2022A&A...668A..74M}). We explored the emission-line ratios involved in the BPT diagram around Mk\,34 (i.e., $\log_{10}\,$([N\textsc{ii}]$\lambda$6583/H$\alpha$)) and $\log_{10}\,$([O\textsc{iii}]$\lambda$5007/H$\beta$)) using Voronoi binning to increase the signal-to-noise ratio.

To create diagnostic diagrams of the region around Mk\,34 we first fitted H$\alpha$ with a single Gaussian in every spaxel of the MUSE cube and used its propagated noise to compute a S/N per pixel. This estimate was then used as input for the Voronoi binning (using the \texttt{vorbin} Python package, \citealt{2003MNRAS.342..345C}), which allowed us to create about $2200$ bins of $S/N\approx500$ in the H$\alpha$ line. To extract the line measurements, the spectra of the resulting bins were then separately fitted over two wavelength ranges (4800$-$5095\AA{}, including H$\beta$ and [O\textsc{iii}]$\lambda\lambda4959,5007$, and He\,I$\lambda5015$, and 6520$-$6994\AA{}, including [N\textsc{ii}]$\lambda\lambda6548,83$, H$\alpha$, He\,I$\lambda6678$, and [S\textsc{ii}]$\lambda\lambda6716,31$) using a model of Gaussian peaks for the emission lines and a seventh-degree polynomial for the continuum. We used the \texttt{pPXF} \citep{2017MNRAS.466..798C} package to carry out the line fitting. The fit leaves residuals in the spatial positions where the light is dominated by stars, especially around the Balmer lines, and we therefore mark these regions. Based on the maps of the line ratios [O\textsc{iii}]$\lambda5007$/H$\beta$ and [N\textsc{ii}]$\lambda6583$/H$\alpha,$ we then manually defined regions of interest using polygon regions in the SAO DS9 viewer \citep{2003ASPC..295..489J}; these are used to color the points in the BPT diagrams. The background-subtracted point in the BPT diagram was created by interpolating over the red region that marks the spatial extent of the outflow. We used AstroPy's \verb|interpolate_replace_nans| function to determine the surrounding background in the map of each line, subtracted it, and then masked all non-positive bins. We then computed the BPT line ratios using the median flux left after subtraction. The error bars are the median absolute deviation of all subtracted bins within the red region.

\begin{figure*}
        \includegraphics[angle=0,width=0.49\textwidth]{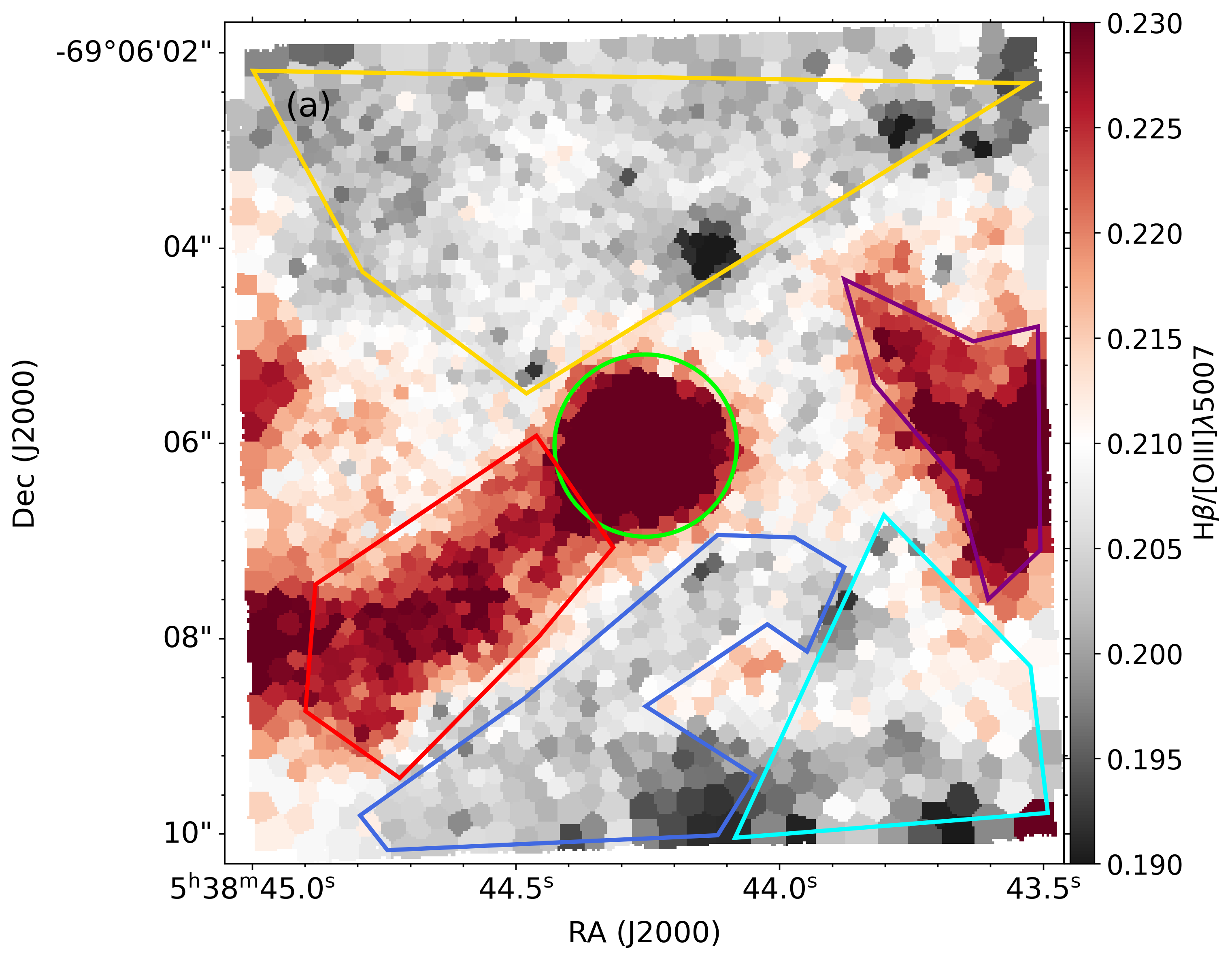}
        \includegraphics[angle=0,width=0.49\textwidth]{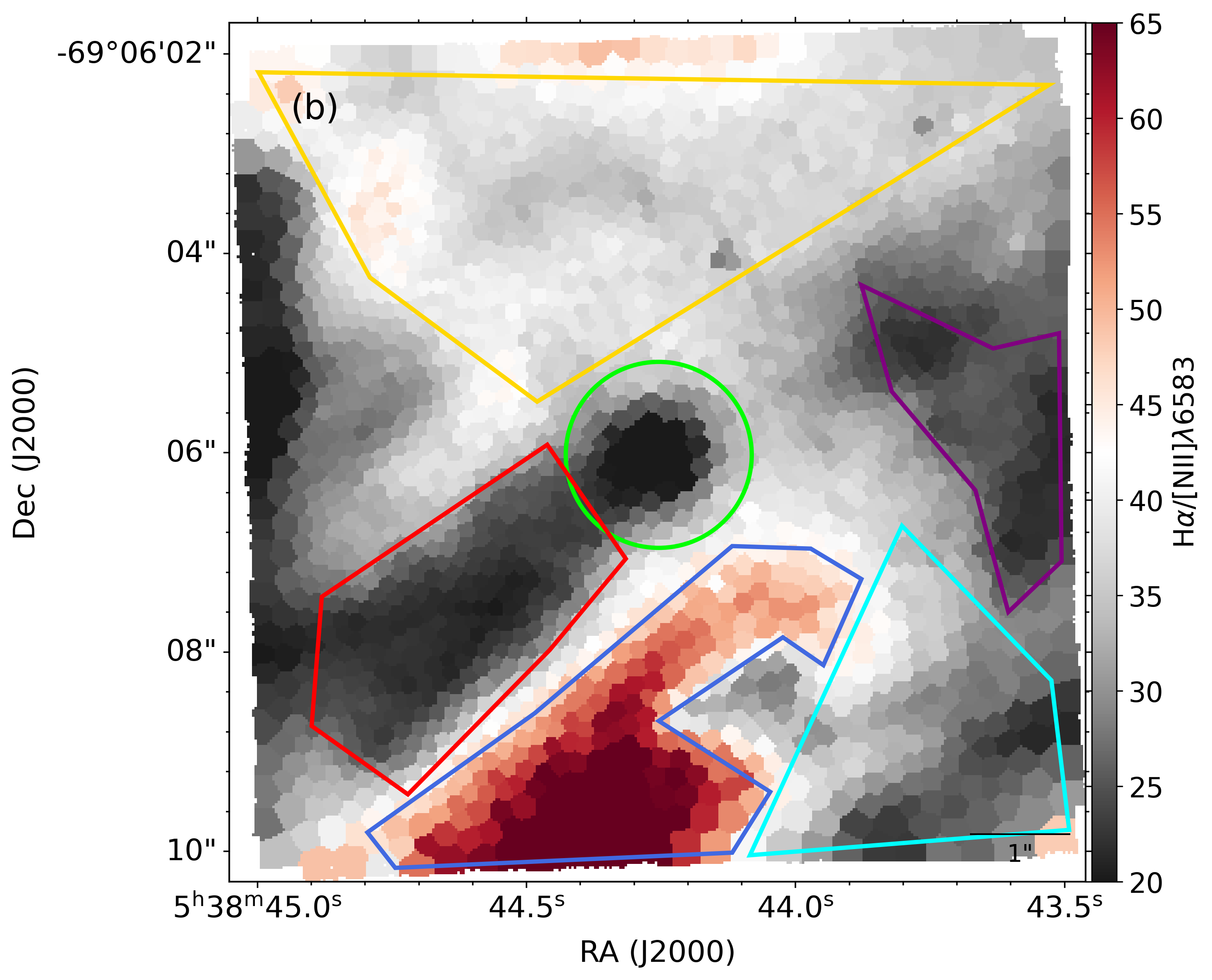}

        \caption{Emission line ratios H$\beta$/[O\textsc{iii}]$\lambda$\,5007 and
 H$\alpha$/[N\textsc{ii}]$\lambda$\,6583  are shown on the left and right panels, respectively. We visually identified several  areas to explore their position and nature in the BPT diagram \citep{1981PASP...93....5B} (colored regions in both panels).    }
        \label{fig:linesRatio}
\end{figure*}

$\log_{10}\,$([N\textsc{ii}]$\lambda$6583/H$\alpha$) and $\log_{10}\,$([O\textsc{iii}]$\lambda$5007/H$\beta$) line ratios (see Fig.~\ref{fig:linesRatio}) highlight the outflow candidate presented in this work. We visually marked several regions in the line-ratio maps and checked their distribution in the BPT diagram. These areas are located in the regions expected to host strong star formation \citep{2017Msngr.170...40C}, matching also the predicted position by numerical BPASS and CLOUDY simluations \citep{2018MNRAS.477..904X}. Additionally, they are located close to the position of the  the so-called green pea galaxies \citep{2009MNRAS.399.1191C}, suggesting similar high-excitation properties. However, we find lower observed electron densities than those suggested by the best-fitting \cite{2018MNRAS.477..904X}'s simulations. This discrepancy in the electron density was also noted by \cite{2018MNRAS.477..904X}, and was attributed in part to the effects of the depletion of metals onto interstellar dust grains.

\begin{figure*}
        \centering
        \includegraphics[angle=0,width=0.49\textwidth]{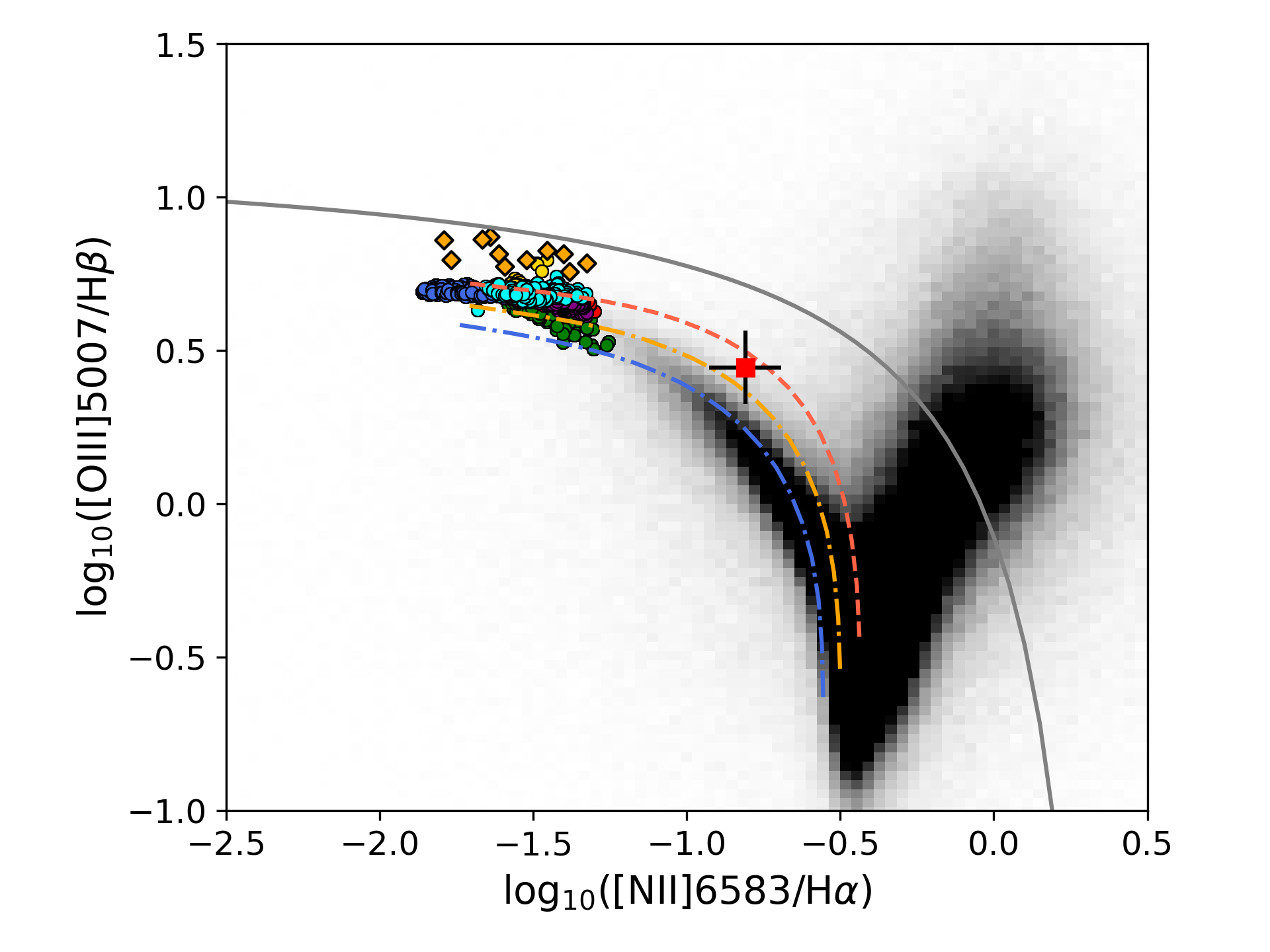}
        \includegraphics[angle=0,width=0.49\textwidth]{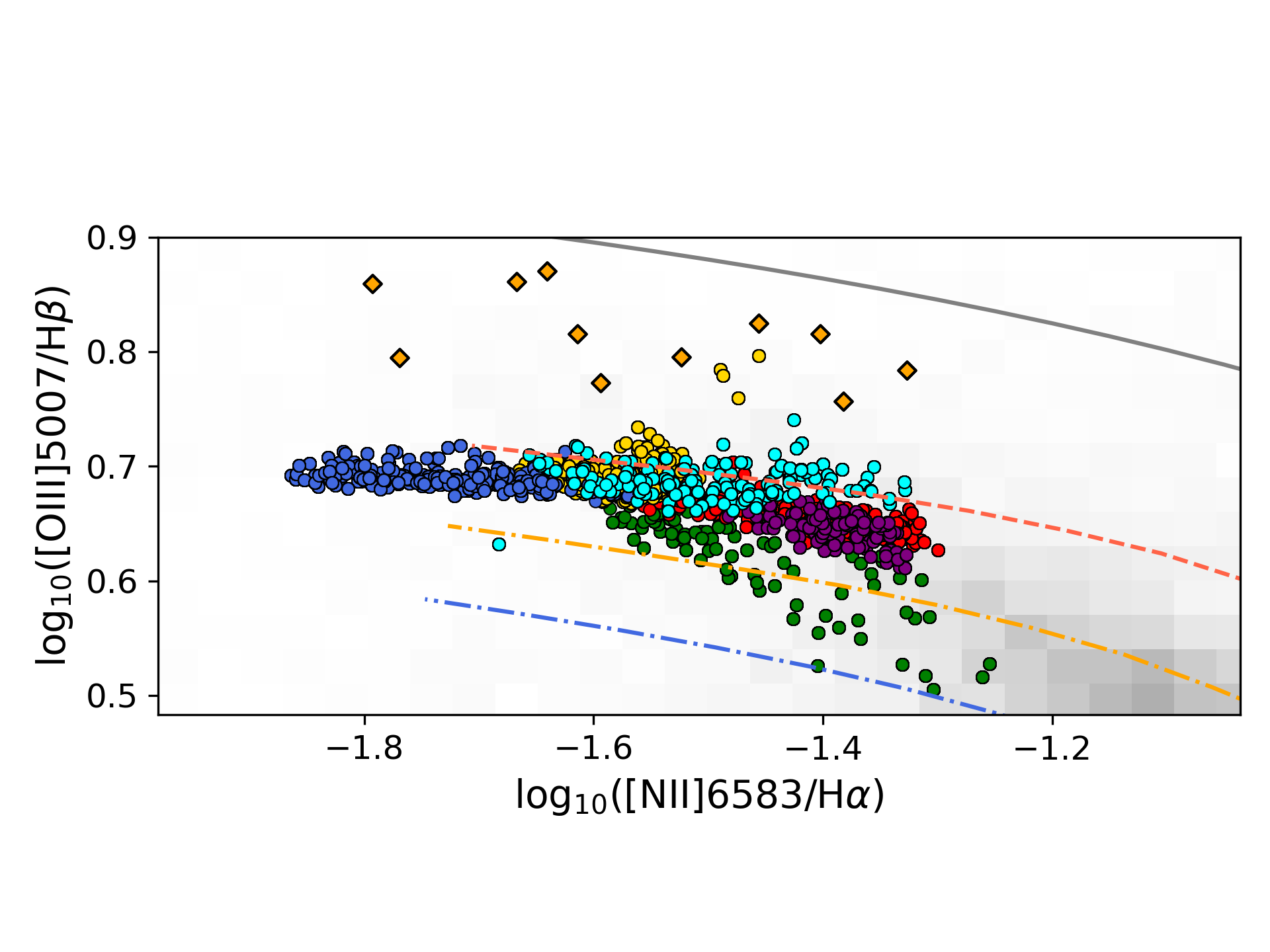}
        
        \caption{Distribution of the areas highlighted in Fig.~\ref{fig:linesRatio} in the BPT diagram \citep{1981PASP...93....5B}. The colored dots follow the color coding used in Fig.~\ref{fig:linesRatio}; the large red point represents the background-subtracted line ratios of the red region. The right panel zooms into the more general view displayed on the left. The BPT diagram extracted from the SDSS DR7 sample \citep{2009ApJS..182..543A} is displayed in the background (gray shaded area) as a reference, together with the \cite{2001ApJ...556..121K} maximum starburst line in both panels. Extreme
Lyman-continuum-emitting green pea galaxies (orange diamonds), updated from
\cite{2017ApJ...845..165M}, are also shown. Additionally, the observed regions are compared with numerical predictions using BPASS and CLOUDY \citep{2018MNRAS.477..904X}. The simulations were calculated using  the Large Magellanic Cloud metallicity, an age of 1\,Myr, and a hydrogen density $\log_{10}\,$(n$_H$[cm$^{-3}$]) ranging between 2 (dotted-dashed blue line), 2.5 (dotted-dashed orange line), and 3 (dashed red line). }
        \label{fig:BPT}
\end{figure*}

Previous studies already highlighted the similarity in nebular properties of 30 Doradus and extreme star-forming  galaxies, including green peas \citep{2017ApJ...845..165M,2023A&A...674A.210M}. Although the regions are not completely separated in the BPT diagram, they are clustering, on average, at slightly different values of $\log_{10}\,$([N\textsc{ii}]$\lambda$6583/H$\alpha$). The outflow (red area in Fig.~\ref{fig:linesRatio}) and the diametrically opposite region (purple area in Fig.~\ref{fig:linesRatio}) are located at similar values in the BPT diagram. Notably, this diametrically opposite region roughly corresponds to the blueshifted area shown in Fig.~\ref{fig:wfm_velo}. The explored areas surrounding Mk\,34 are systematically located below the average values reported in the extreme Lyman-continuum-emitting green pea galaxies \citep{2017ApJ...845..165M} in the BPT diagram (Fig.~\ref{fig:BPT}).
Because the outflow consists of ionized gas that adds about 10\% flux to the H$\alpha$ surface brightness, we also aimed to subtract the surrounding ionized background to isolate the line ratios due to the outflow itself. This resulting background-corrected median flux is located in a very different region of the BPT diagram (red point in Fig.~\ref{fig:BPT}) where shock models are located. Specifically, the MAPPINGS III models \citep{2008ApJS..178...20A}, which include shock and precursor, cover exactly this area for LMC metallicity, $200\lesssim{}V\lesssim{}250\,$km/s, and $1\lesssim{}B\lesssim{}5\,\mu{}$G. We see this as an indication that the outflow from Mk\,34 induces shocks in the surrounding ISM.

\section{Discussion and future steps }
\label{conclusion}

Reported outflows in  massive objects are scarce and are mainly detected at radio wavelengths \citep{2010Sci...330.1209C}. Several detections have also been reported at optical and IR wavelengths in stellar nurseries, where stellar sources are still deeply buried in the molecular cloud \citep{2015A&A...573A..82C,2018Natur.554..334M}. Mk\,34 is not a star in the formation phase. It is a system of two-core hydrogen-burning  very massive stars, where the nursery cloud was swept away some time ago. The fact that the stellar sources and outflow can be studied separately makes this system an excellent target with which to empirically explore the origin of these ejections into the ISM. The colliding stellar winds in very massive systems can indeed generate outflows and eject processed material further into the ISM \citep{1990FlDy...25..629L,2012A&A...546A..60L}. The high spatial resolution of MUSE-NFM confirms the impact of outflow in the ISM, opening channels for the ionizing radiation to escape \citep{2020A&A...644A..21R,2020AJ....160...78R}, as shown in Fig.~\ref{fig:wfm}.

The maps defined by the [N\textsc{ii}] and [S\textsc{ii}] emission lines reveal a cone-like structure originating from Mk\,34, suggesting a plausible outflow. This finding is also confirmed by the electron density and ISM radial velocity maps around Mk\,34. The ISM material appears to propagate toward the southeast region of Mk\,34, with no clear counterpart observed in the northwest of the system. However, along the direction marked by the [N\textsc{ii}] and [S\textsc{ii}] filaments, the material exhibits similar properties on both sides of Mk\,34, as indicated by the emission-line ratios in the BPT diagram. We detect greater extinction on the northwest side of Mk\,34, which correlates with the lack of [N\textsc{ii}] emission and may explain the absence of a corresponding outflow in that direction.

 These ISM characteristics were consistently observed across both epochs. Between the 2019 and 2021 observations, we noted changes primarily in the intensity maps, as well as in the electron density and extinction within a 0.5" radius of Mk\,34, indicating possible monthly variations in the system. Transient phenomena, such as the formation and dissipation of emission disks, are not uncommon and have been documented in several Oe/Be stars \citep{2013A&ARv..21...69R}. A notable example is AzV493, a scarce Oe-type star in the Small Magellanic Cloud (SMC), which exhibits significant spectroscopic variations dominated by a 14.6\,year cycle, along with intense photometric fluctuations on shorter timescales (months) and approximately 40\,day oscillations \citep{2023arXiv230111433O}. The variability pattern observed in Mk\,34 aligns with the expected behavior of massive, dynamically active systems of this kind.

We modeled and subtracted the Mk\,34 MUSE-NFM PSF to search for additional signals close to the star. The 2019 PSF residual map reveals several positive regions near Mk\,34 within a circle of 0.5" in radius centered on the system. This pattern is consistently observed in the individual analysis of each wavelength, providing robust results. The possibility that we are witnessing bursts originating from the system is intriguing and not entirely unexpected. The collision of the supersonic stellar winds produced by the twin WRs in Mk\,34 could result in rapid outflows \citep{1990FlDy...25..629L,2012A&A...546A..60L}. However, we must exercise caution and seek additional observations to further investigate this phenomenon. The change in the pattern observed in 2021 once again highlights the need for further observations in future epochs.

MUSE-NFM has the capability of spectroscopically mapping the impact of the feedback  from the most massive stars on the surrounding ISM, and could be used to examine the contribution of this feedback to the ionization and chemical composition of the cluster. The present work conducted on Mk\,34 is a demonstration of these capabilities,  with MUSE-NFM achieving a spatial resolution of $\approx$\,80\,mas in both epochs. The detection of the [N\textsc{ii}] and [S\textsc{ii}] filaments originating from Mk\,34 suggests we that it may be possible to develop a better observational characterization of the dense and extensive stellar winds of Mk\,34 and binary interactions.  Numerical radiation-hydrodynamic modeling of the stellar winds of  massive stars shows that time-dependent wind develops a very inhomogeneous clumped structure \citep{1988ApJ...335..914O,2018A&A...611A..17S}. The current state-of-the-art high-spatial-resolution spectrographs are excellent instruments  with which to explore the extended stellar winds of the most massive stars known and to record their development over  time. Comprehensively understanding dynamic events such as those observed in Mk\,34 requires supplementary data that enable continuous monitoring. Additional observations of Mk\,34 will be requested to investigate short-term variations that cannot be adequately mapped with the two epochs separated by 745 days in the present work, and to explore a potential correlation with the 155\,day period of the Mk\,34 binary system.

\begin{acknowledgements}

The authors thanks the referee for useful comments and helpful suggestions that improved this manuscript. NC acknowledges funding from the Deutsche Forschungsgemeinschaft (DFG) - CA 2551/1-1,2 and BMBF ErUM (project VLTBlueMUSE, grant 05A23MGA). PMW received support from BMBF ErUM (VLT-BlueMUSE and BlueMUSE-CD, grants 05A20BAB and 05A23BAC). Our research used Astropy, a community-developed core Pythonpackage for Astronomy \citep{2013A&A...558A..33A}, and APLpy, an open-source plotting package for Python \citep{2012ascl.soft08017R}. 
\end{acknowledgements}

\bibliographystyle{aa}
\bibliography{Mk34_jets}

\begin{thebibliography}{57}
\expandafter\ifx\csname natexlab\endcsname\relax\def\natexlab#1{#1}\fi

\bibitem[{{Abazajian} {et~al.}(2009){Abazajian}, {Adelman-McCarthy},
  {Ag{\"u}eros}, {Allam}, {Allende Prieto}, {An}, {Anderson}, {Anderson},
  {Annis}, {Bahcall}, {Bailer-Jones}, {Barentine}, {Bassett}, {Becker},
  {Beers}, {Bell}, {Belokurov}, {Berlind}, {Berman}, {Bernardi}, {Bickerton},
  {Bizyaev}, {Blakeslee}, {Blanton}, {Bochanski}, {Boroski}, {Brewington},
  {Brinchmann}, {Brinkmann}, {Brunner}, {Budav{\'a}ri}, {Carey}, {Carliles},
  {Carr}, {Castander}, {Cinabro}, {Connolly}, {Csabai}, {Cunha}, {Czarapata},
  {Davenport}, {de Haas}, {Dilday}, {Doi}, {Eisenstein}, {Evans}, {Evans},
  {Fan}, {Friedman}, {Frieman}, {Fukugita}, {G{\"a}nsicke}, {Gates},
  {Gillespie}, {Gilmore}, {Gonzalez}, {Gonzalez}, {Grebel}, {Gunn},
  {Gy{\"o}ry}, {Hall}, {Harding}, {Harris}, {Harvanek}, {Hawley}, {Hayes},
  {Heckman}, {Hendry}, {Hennessy}, {Hindsley}, {Hoblitt}, {Hogan}, {Hogg},
  {Holtzman}, {Hyde}, {Ichikawa}, {Ichikawa}, {Im}, {Ivezi{\'c}}, {Jester},
  {Jiang}, {Johnson}, {Jorgensen}, {Juri{\'c}}, {Kent}, {Kessler}, {Kleinman},
  {Knapp}, {Konishi}, {Kron}, {Krzesinski}, {Kuropatkin}, {Lampeitl},
  {Lebedeva}, {Lee}, {Lee}, {French Leger}, {L{\'e}pine}, {Li}, {Lima}, {Lin},
  {Long}, {Loomis}, {Loveday}, {Lupton}, {Magnier}, {Malanushenko},
  {Malanushenko}, {Mandelbaum}, {Margon}, {Marriner}, {Mart{\'\i}nez-Delgado},
  {Matsubara}, {McGehee}, {McKay}, {Meiksin}, {Morrison}, {Mullally}, {Munn},
  {Murphy}, {Nash}, {Nebot}, {Neilsen}, {Newberg}, {Newman}, {Nichol},
  {Nicinski}, {Nieto-Santisteban}, {Nitta}, {Okamura}, {Oravetz}, {Ostriker},
  {Owen}, {Padmanabhan}, {Pan}, {Park}, {Pauls}, {Peoples}, {Percival}, {Pier},
  {Pope}, {Pourbaix}, {Price}, {Purger}, {Quinn}, {Raddick}, {Re Fiorentin},
  {Richards}, {Richmond}, {Riess}, {Rix}, {Rockosi}, {Sako}, {Schlegel},
  {Schneider}, {Scholz}, {Schreiber}, {Schwope}, {Seljak}, {Sesar}, {Sheldon},
  {Shimasaku}, {Sibley}, {Simmons}, {Sivarani}, {Allyn Smith}, {Smith},
  {Smol{\v{c}}i{\'c}}, {Snedden}, {Stebbins}, {Steinmetz}, {Stoughton},
  {Strauss}, {SubbaRao}, {Suto}, {Szalay}, {Szapudi}, {Szkody}, {Tanaka},
  {Tegmark}, {Teodoro}, {Thakar}, {Tremonti}, {Tucker}, {Uomoto}, {Vanden
  Berk}, {Vandenberg}, {Vidrih}, {Vogeley}, {Voges}, {Vogt}, {Wadadekar},
  {Watters}, {Weinberg}, {West}, {White}, {Wilhite}, {Wonders}, {Yanny},
  {Yocum}, {York}, {Zehavi}, {Zibetti}, \& {Zucker}}]{2009ApJS..182..543A}
{Abazajian}, K.~N., {Adelman-McCarthy}, J.~K., {Ag{\"u}eros}, M.~A., {et~al.}
  2009, \apjs, 182, 543

\bibitem[{{Allen} {et~al.}(2008){Allen}, {Groves}, {Dopita}, {Sutherland}, \&
  {Kewley}}]{2008ApJS..178...20A}
{Allen}, M.~G., {Groves}, B.~A., {Dopita}, M.~A., {Sutherland}, R.~S., \&
  {Kewley}, L.~J. 2008, \apjs, 178, 20

\bibitem[{{Astropy Collaboration} {et~al.}(2013){Astropy Collaboration},
  {Robitaille}, {Tollerud}, {Greenfield}, {Droettboom}, {Bray}, {Aldcroft},
  {Davis}, {Ginsburg}, {Price-Whelan}, {Kerzendorf}, {Conley}, {Crighton},
  {Barbary}, {Muna}, {Ferguson}, {Grollier}, {Parikh}, {Nair}, {Unther},
  {Deil}, {Woillez}, {Conseil}, {Kramer}, {Turner}, {Singer}, {Fox}, {Weaver},
  {Zabalza}, {Edwards}, {Azalee Bostroem}, {Burke}, {Casey}, {Crawford},
  {Dencheva}, {Ely}, {Jenness}, {Labrie}, {Lim}, {Pierfederici}, {Pontzen},
  {Ptak}, {Refsdal}, {Servillat}, \& {Streicher}}]{2013A&A...558A..33A}
{Astropy Collaboration}, {Robitaille}, T.~P., {Tollerud}, E.~J., {et~al.} 2013,
  \aap, 558, A33

\bibitem[{{Bacon} {et~al.}(2010){Bacon}, {Accardo}, {Adjali}, {Anwand},
  {Bauer}, {Biswas}, {Blaizot}, {Boudon}, {Brau-Nogue}, {Brinchmann},
  {Caillier}, {Capoani}, {Carollo}, {Contini}, {Couderc}, {Daguis{\'e}},
  {Deiries}, {Delabre}, {Dreizler}, {Dubois}, {Dupieux}, {Dupuy}, {Emsellem},
  {Fechner}, {Fleischmann}, {Fran{\c{c}}ois}, {Gallou}, {Gharsa}, {Glindemann},
  {Gojak}, {Guiderdoni}, {Hansali}, {Hahn}, {Jarno}, {Kelz}, {Koehler},
  {Kosmalski}, {Laurent}, {Le Floch}, {Lilly}, {Lizon}, {Loupias}, {Manescau},
  {Monstein}, {Nicklas}, {Olaya}, {Pares}, {Pasquini}, {P{\'e}contal-Rousset},
  {Pell{\'o}}, {Petit}, {Popow}, {Reiss}, {Remillieux}, {Renault}, {Roth},
  {Rupprecht}, {Serre}, {Schaye}, {Soucail}, {Steinmetz}, {Streicher}, {Stuik},
  {Valentin}, {Vernet}, {Weilbacher}, {Wisotzki}, \&
  {Yerle}}]{2010SPIE.7735E..08B}
{Bacon}, R., {Accardo}, M., {Adjali}, L., {et~al.} 2010, in Society of
  Photo-Optical Instrumentation Engineers (SPIE) Conference Series, Vol. 7735,
  Ground-based and Airborne Instrumentation for Astronomy III, ed. I.~S.
  {McLean}, S.~K. {Ramsay}, \& H.~{Takami}, 773508

\bibitem[{{Bacon} {et~al.}(2014){Bacon}, {Vernet}, {Borisova}, {Bouch{\'e}},
  {Brinchmann}, {Carollo}, {Carton}, {Caruana}, {Cerda}, {Contini}, {Franx},
  {Girard}, {Guerou}, {Haddad}, {Hau}, {Herenz}, {Herrera}, {Husemann},
  {Husser}, {Jarno}, {Kamann}, {Krajnovic}, {Lilly}, {Mainieri}, {Martinsson},
  {Palsa}, {Patricio}, {P{\'e}contal}, {Pello}, {Piqueras}, {Richard},
  {Sandin}, {Schroetter}, {Selman}, {Shirazi}, {Smette}, {Soto}, {Streicher},
  {Urrutia}, {Weilbacher}, {Wisotzki}, \& {Zins}}]{2014Msngr.157...13B}
{Bacon}, R., {Vernet}, J., {Borisova}, E., {et~al.} 2014, The Messenger, 157,
  13

\bibitem[{{Baldwin} {et~al.}(1981){Baldwin}, {Phillips}, \&
  {Terlevich}}]{1981PASP...93....5B}
{Baldwin}, J.~A., {Phillips}, M.~M., \& {Terlevich}, R. 1981, \pasp, 93, 5

\bibitem[{{Belczynski} {et~al.}(2022){Belczynski}, {Romagnolo}, {Olejak},
  {Klencki}, {Chattopadhyay}, {Stevenson}, {Coleman Miller}, {Lasota}, \&
  {Crowther}}]{2022ApJ...925...69B}
{Belczynski}, K., {Romagnolo}, A., {Olejak}, A., {et~al.} 2022, \apj, 925, 69

\bibitem[{{Cappellari}(2017)}]{2017MNRAS.466..798C}
{Cappellari}, M. 2017, \mnras, 466, 798

\bibitem[{{Cappellari} \& {Copin}(2003)}]{2003MNRAS.342..345C}
{Cappellari}, M. \& {Copin}, Y. 2003, \mnras, 342, 345

\bibitem[{{Caratti o Garatti} {et~al.}(2015){Caratti o Garatti}, {Stecklum},
  {Linz}, {Garcia Lopez}, \& {Sanna}}]{2015A&A...573A..82C}
{Caratti o Garatti}, A., {Stecklum}, B., {Linz}, H., {Garcia Lopez}, R., \&
  {Sanna}, A. 2015, \aap, 573, A82

\bibitem[{{Cardamone} {et~al.}(2009){Cardamone}, {Schawinski}, {Sarzi},
  {Bamford}, {Bennert}, {Urry}, {Lintott}, {Keel}, {Parejko}, {Nichol},
  {Thomas}, {Andreescu}, {Murray}, {Raddick}, {Slosar}, {Szalay}, \&
  {Vandenberg}}]{2009MNRAS.399.1191C}
{Cardamone}, C., {Schawinski}, K., {Sarzi}, M., {et~al.} 2009, \mnras, 399,
  1191

\bibitem[{{Cardelli} {et~al.}(1989){Cardelli}, {Clayton}, \&
  {Mathis}}]{1989ApJ...345..245C}
{Cardelli}, J.~A., {Clayton}, G.~C., \& {Mathis}, J.~S. 1989, \apj, 345, 245

\bibitem[{{Carrasco-Gonz{\'a}lez} {et~al.}(2010){Carrasco-Gonz{\'a}lez},
  {Rodr{\'\i}guez}, {Anglada}, {Mart{\'\i}}, {Torrelles}, \&
  {Osorio}}]{2010Sci...330.1209C}
{Carrasco-Gonz{\'a}lez}, C., {Rodr{\'\i}guez}, L.~F., {Anglada}, G., {et~al.}
  2010, Science, 330, 1209

\bibitem[{{Castro} {et~al.}(2018){Castro}, {Crowther}, {Evans}, {Mackey},
  {Castro-Rodriguez}, {Vink}, {Melnick}, \& {Selman}}]{2018AA...614A.147C}
{Castro}, N., {Crowther}, P.~A., {Evans}, C.~J., {et~al.} 2018, \aap, 614, A147

\bibitem[{{Castro} {et~al.}(2021{\natexlab{a}}){Castro}, {Crowther}, {Evans},
  {Vink}, {Puls}, {Herrero}, {Garcia}, {Selman}, {Roth}, \&
  {Sim{\'o}n-D{\'\i}az}}]{2021A&A...648A..65C}
{Castro}, N., {Crowther}, P.~A., {Evans}, C.~J., {et~al.} 2021{\natexlab{a}},
  \aap, 648, A65

\bibitem[{{Castro} {et~al.}(2021{\natexlab{b}}){Castro}, {Roth}, {Weilbacher},
  {Micheva}, {Monreal-Ibero}, {Kelz}, {Kamann}, {Maseda}, {Wendt}, \& {MUSE
  Collaboration}}]{2021Msngr.182...50C}
{Castro}, N., {Roth}, M.~M., {Weilbacher}, P.~M., {et~al.} 2021{\natexlab{b}},
  The Messenger, 182, 50

\bibitem[{{Chen{\'e}} {et~al.}(2011){Chen{\'e}}, {Schnurr}, {Crowther},
  {Fern{\'a}ndez-Laj{\'u}s}, \& {Moffat}}]{2011IAUS..272..497C}
{Chen{\'e}}, A.-N., {Schnurr}, O., {Crowther}, P.~A.,
  {Fern{\'a}ndez-Laj{\'u}s}, E., \& {Moffat}, A. F.~J. 2011, in Active OB
  Stars: Structure, Evolution, Mass Loss, and Critical Limits, ed. C.~{Neiner},
  G.~{Wade}, G.~{Meynet}, \& G.~{Peters}, Vol. 272, 497--498

\bibitem[{{Crowther} {et~al.}(2022){Crowther}, {Broos}, {Townsley}, {Pollock},
  {Tehrani}, \& {Gagn{\'e}}}]{2022MNRAS.515.4130C}
{Crowther}, P.~A., {Broos}, P.~S., {Townsley}, L.~K., {et~al.} 2022, \mnras,
  515, 4130

\bibitem[{{Crowther} {et~al.}(2017){Crowther}, {Castro}, {Evans}, {Vink},
  {Melnick}, \& {Selman}}]{2017Msngr.170...40C}
{Crowther}, P.~A., {Castro}, N., {Evans}, C.~J., {et~al.} 2017, The Messenger,
  170, 40

\bibitem[{{Crowther} \& {Dessart}(1998)}]{1998MNRAS.296..622C}
{Crowther}, P.~A. \& {Dessart}, L. 1998, \mnras, 296, 622

\bibitem[{{Crowther} {et~al.}(2010){Crowther}, {Schnurr}, {Hirschi}, {Yusof},
  {Parker}, {Goodwin}, \& {Kassim}}]{2010MNRAS.408..731C}
{Crowther}, P.~A., {Schnurr}, O., {Hirschi}, R., {et~al.} 2010, \mnras, 408,
  731

\bibitem[{{Crowther} \& {Walborn}(2011)}]{2011MNRAS.416.1311C}
{Crowther}, P.~A. \& {Walborn}, N.~R. 2011, \mnras, 416, 1311

\bibitem[{{Doran} {et~al.}(2013){Doran}, {Crowther}, {de Koter}, {Evans},
  {McEvoy}, {Walborn}, {Bastian}, {Bestenlehner}, {Gr{\"a}fener}, {Herrero},
  {K{\"o}hler}, {Ma{\'\i}z Apell{\'a}niz}, {Najarro}, {Puls}, {Sana},
  {Schneider}, {Taylor}, {van Loon}, \& {Vink}}]{2013A&A...558A.134D}
{Doran}, E.~I., {Crowther}, P.~A., {de Koter}, A., {et~al.} 2013, \aap, 558,
  A134

\bibitem[{{Eldridge} \& {Stanway}(2022)}]{2022arXiv220201413E}
{Eldridge}, J.~J. \& {Stanway}, E.~R. 2022, arXiv e-prints, arXiv:2202.01413

\bibitem[{{F{\'e}tick} {et~al.}(2019){F{\'e}tick}, {Fusco}, {Neichel},
  {Mugnier}, {Beltramo-Martin}, {Bonnefois}, {Petit}, {Milli}, {Vernet},
  {Oberti}, \& {Bacon}}]{2019A&A...628A..99F}
{F{\'e}tick}, R.~J.~L., {Fusco}, T., {Neichel}, B., {et~al.} 2019, \aap, 628,
  A99

\bibitem[{{Hainich} {et~al.}(2014){Hainich}, {R{\"u}hling}, {Todt}, {Oskinova},
  {Liermann}, {Gr{\"a}fener}, {Foellmi}, {Schnurr}, \&
  {Hamann}}]{2014A&A...565A..27H}
{Hainich}, R., {R{\"u}hling}, U., {Todt}, H., {et~al.} 2014, \aap, 565, A27

\bibitem[{{Hummer} \& {Storey}(1987)}]{1987MNRAS.224..801H}
{Hummer}, D.~G. \& {Storey}, P.~J. 1987, \mnras, 224, 801

\bibitem[{{Joye} \& {Mandel}(2003)}]{2003ASPC..295..489J}
{Joye}, W.~A. \& {Mandel}, E. 2003, in Astronomical Society of the Pacific
  Conference Series, Vol. 295, Astronomical Data Analysis Software and Systems
  XII, ed. H.~E. {Payne}, R.~I. {Jedrzejewski}, \& R.~N. {Hook}, 489

\bibitem[{{Kewley} {et~al.}(2001){Kewley}, {Dopita}, {Sutherland}, {Heisler},
  \& {Trevena}}]{2001ApJ...556..121K}
{Kewley}, L.~J., {Dopita}, M.~A., {Sutherland}, R.~S., {Heisler}, C.~A., \&
  {Trevena}, J. 2001, \apj, 556, 121

\bibitem[{{Lamberts} {et~al.}(2012){Lamberts}, {Dubus}, {Lesur}, \&
  {Fromang}}]{2012A&A...546A..60L}
{Lamberts}, A., {Dubus}, G., {Lesur}, G., \& {Fromang}, S. 2012, \aap, 546, A60

\bibitem[{{Lebedev} \& {Myasnikov}(1990)}]{1990FlDy...25..629L}
{Lebedev}, M.~G. \& {Myasnikov}, A.~V. 1990, Fluid Dynamics, 25, 629

\bibitem[{{Lindegren} {et~al.}(2018){Lindegren}, {Hern{\'a}ndez}, {Bombrun},
  {Klioner}, {Bastian}, {Ramos-Lerate}, {de Torres}, {Steidelm{\"u}ller},
  {Stephenson}, {Hobbs}, {Lammers}, {Biermann}, {Geyer}, {Hilger}, {Michalik},
  {Stampa}, {McMillan}, {Casta{\~n}eda}, {Clotet}, {Comoretto}, {Davidson},
  {Fabricius}, {Gracia}, {Hambly}, {Hutton}, {Mora}, {Portell}, {van Leeuwen},
  {Abbas}, {Abreu}, {Altmann}, {Andrei}, {Anglada}, {Balaguer-N{\'u}{\~n}ez},
  {Barache}, {Becciani}, {Bertone}, {Bianchi}, {Bouquillon}, {Bourda},
  {Br{\"u}semeister}, {Bucciarelli}, {Busonero}, {Buzzi}, {Cancelliere},
  {Carlucci}, {Charlot}, {Cheek}, {Crosta}, {Crowley}, {de Bruijne}, {de
  Felice}, {Drimmel}, {Esquej}, {Fienga}, {Fraile}, {Gai}, {Garralda},
  {Gonz{\'a}lez-Vidal}, {Guerra}, {Hauser}, {Hofmann}, {Holl}, {Jordan},
  {Lattanzi}, {Lenhardt}, {Liao}, {Licata}, {Lister}, {L{\"o}ffler},
  {Marchant}, {Martin-Fleitas}, {Messineo}, {Mignard}, {Morbidelli}, {Poggio},
  {Riva}, {Rowell}, {Salguero}, {Sarasso}, {Sciacca}, {Siddiqui}, {Smart},
  {Spagna}, {Steele}, {Taris}, {Torra}, {van Elteren}, {van Reeven}, \&
  {Vecchiato}}]{GaiaDR2_astrometry}
{Lindegren}, L., {Hern{\'a}ndez}, J., {Bombrun}, A., {et~al.} 2018, \aap, 616,
  A2

\bibitem[{{McCall}(1984)}]{1984MNRAS.208..253M}
{McCall}, M.~L. 1984, \mnras, 208, 253

\bibitem[{{McLeod} {et~al.}(2018){McLeod}, {Reiter}, {Kuiper}, {Klaassen}, \&
  {Evans}}]{2018Natur.554..334M}
{McLeod}, A.~F., {Reiter}, M., {Kuiper}, R., {Klaassen}, P.~D., \& {Evans},
  C.~J. 2018, \nat, 554, 334

\bibitem[{{Micheva} {et~al.}(2019){Micheva}, {Christian Herenz}, {Roth},
  {{\"O}stlin}, \& {Girichidis}}]{2019A&A...623A.145M}
{Micheva}, G., {Christian Herenz}, E., {Roth}, M.~M., {{\"O}stlin}, G., \&
  {Girichidis}, P. 2019, \aap, 623, A145

\bibitem[{{Micheva} {et~al.}(2017){Micheva}, {Oey}, {Jaskot}, \&
  {James}}]{2017ApJ...845..165M}
{Micheva}, G., {Oey}, M.~S., {Jaskot}, A.~E., \& {James}, B.~L. 2017, \aj, 845,
  165

\bibitem[{{Micheva} {et~al.}(2022){Micheva}, {Roth}, {Weilbacher}, {Morisset},
  {Castro}, {Monreal Ibero}, {Soemitro}, {Maseda}, {Steinmetz}, \&
  {Brinchmann}}]{2022A&A...668A..74M}
{Micheva}, G., {Roth}, M.~M., {Weilbacher}, P.~M., {et~al.} 2022, \aap, 668,
  A74

\bibitem[{{Monreal-Ibero} {et~al.}(2023){Monreal-Ibero}, {Weilbacher},
  {Micheva}, {Kollatschny}, \& {Maseda}}]{2023A&A...674A.210M}
{Monreal-Ibero}, A., {Weilbacher}, P.~M., {Micheva}, G., {Kollatschny}, W., \&
  {Maseda}, M. 2023, \aap, 674, A210

\bibitem[{{Oey} {et~al.}(2023){Oey}, {Castro}, {Renzo}, {Vargas-Salazar},
  {Suffak}, {Ratajczak}, {Monnier}, {Szymanski}, {Phillips}, {Calvet}, {Chiti},
  {Micheva}, {Rasmussen}, \& {Townsend}}]{2023arXiv230111433O}
{Oey}, M.~S., {Castro}, N., {Renzo}, M., {et~al.} 2023, arXiv e-prints,
  arXiv:2301.11433

\bibitem[{{Owocki} {et~al.}(1988){Owocki}, {Castor}, \&
  {Rybicki}}]{1988ApJ...335..914O}
{Owocki}, S.~P., {Castor}, J.~I., \& {Rybicki}, G.~B. 1988, \apj, 335, 914

\bibitem[{{Pietrzy{\'n}ski} {et~al.}(2013){Pietrzy{\'n}ski}, {Graczyk},
  {Gieren}, {Thompson}, {Pilecki}, {Udalski}, {Soszy{\'n}ski}, {Koz{\l}owski},
  {Konorski}, {Suchomska}, {Bono}, {Moroni}, {Villanova}, {Nardetto},
  {Bresolin}, {Kudritzki}, {Storm}, {Gallenne}, {Smolec}, {Minniti}, {Kubiak},
  {Szyma{\'n}ski}, {Poleski}, {Wyrzykowski}, {Ulaczyk}, {Pietrukowicz},
  {G{\'o}rski}, \& {Karczmarek}}]{2013Natur.495...76P}
{Pietrzy{\'n}ski}, G., {Graczyk}, D., {Gieren}, W., {et~al.} 2013, \nat, 495,
  76

\bibitem[{{Pollock} {et~al.}(2018){Pollock}, {Crowther}, {Tehrani}, {Broos}, \&
  {Townsley}}]{2018MNRAS.474.3228P}
{Pollock}, A.~M.~T., {Crowther}, P.~A., {Tehrani}, K., {Broos}, P.~S., \&
  {Townsley}, L.~K. 2018, \mnras, 474, 3228

\bibitem[{{Ramambason} {et~al.}(2020){Ramambason}, {Schaerer}, {Stasi{\'n}ska},
  {Izotov}, {Guseva}, {V{\'\i}lchez}, {Amor{\'\i}n}, \&
  {Morisset}}]{2020A&A...644A..21R}
{Ramambason}, L., {Schaerer}, D., {Stasi{\'n}ska}, G., {et~al.} 2020, \aap,
  644, A21

\bibitem[{{Rivinius} {et~al.}(2013){Rivinius}, {Carciofi}, \&
  {Martayan}}]{2013A&ARv..21...69R}
{Rivinius}, T., {Carciofi}, A.~C., \& {Martayan}, C. 2013, \aapr, 21, 69

\bibitem[{{Robitaille} \& {Bressert}(2012)}]{2012ascl.soft08017R}
{Robitaille}, T. \& {Bressert}, E. 2012, {APLpy: Astronomical Plotting Library
  in Python}, Astrophysics Source Code Library, record ascl:1208.017

\bibitem[{{Rosen} \& {Krumholz}(2020)}]{2020AJ....160...78R}
{Rosen}, A.~L. \& {Krumholz}, M.~R. 2020, \aj, 160, 78

\bibitem[{{Roth} {et~al.}(2019){Roth}, {Weilbacher}, \&
  {Castro}}]{2019AN....340..989R}
{Roth}, M.~M., {Weilbacher}, P.~M., \& {Castro}, N. 2019, Astronomische
  Nachrichten, 340, 989

\bibitem[{{Schnurr} {et~al.}(2008){Schnurr}, {Moffat}, {St-Louis}, {Morrell},
  \& {Guerrero}}]{2008MNRAS.389..806S}
{Schnurr}, O., {Moffat}, A.~F.~J., {St-Louis}, N., {Morrell}, N.~I., \&
  {Guerrero}, M.~A. 2008, \mnras, 389, 806

\bibitem[{{Sundqvist} {et~al.}(2018){Sundqvist}, {Owocki}, \&
  {Puls}}]{2018A&A...611A..17S}
{Sundqvist}, J.~O., {Owocki}, S.~P., \& {Puls}, J. 2018, \aap, 611, A17

\bibitem[{{Tehrani} {et~al.}(2019){Tehrani}, {Crowther}, {Bestenlehner},
  {Littlefair}, {Pollock}, {Parker}, \& {Schnurr}}]{2019MNRAS.484.2692T}
{Tehrani}, K.~A., {Crowther}, P.~A., {Bestenlehner}, J.~M., {et~al.} 2019,
  \mnras, 484, 2692

\bibitem[{{Tody}(1993)}]{1993ASPC...52..173T}
{Tody}, D. 1993, in Astronomical Society of the Pacific Conference Series,
  Vol.~52, Astronomical Data Analysis Software and Systems II, ed. R.~J.
  {Hanisch}, R.~J.~V. {Brissenden}, \& J.~{Barnes}, 173

\bibitem[{{Townsley} {et~al.}(2006){Townsley}, {Broos}, {Feigelson}, {Garmire},
  \& {Getman}}]{2006AJ....131.2164T}
{Townsley}, L.~K., {Broos}, P.~S., {Feigelson}, E.~D., {Garmire}, G.~P., \&
  {Getman}, K.~V. 2006, \aj, 131, 2164

\bibitem[{{Vink} {et~al.}(2015){Vink}, {Heger}, {Krumholz}, {Puls}, {Banerjee},
  {Castro}, {Chen}, {Chen{\`e}}, {Crowther}, {Daminelli}, {Gr{\"a}fener},
  {Groh}, {Hamann}, {Heap}, {Herrero}, {Kaper}, {Najarro}, {Oskinova},
  {Roman-Lopes}, {Rosen}, {Sander}, {Shirazi}, {Sugawara}, {Tramper},
  {Vanbeveren}, {Voss}, {Wofford}, \& {Zhang}}]{2015HiA....16...51V}
{Vink}, J.~S., {Heger}, A., {Krumholz}, M.~R., {et~al.} 2015, Highlights of
  Astronomy, 16, 51

\bibitem[{{Vriend}(2015)}]{musewise}
{Vriend}, W.-J. 2015, in Science Operations 2015: Science Data Management - An
  ESO/ESA Workshop, 1

\bibitem[{{Weilbacher} {et~al.}(2018){Weilbacher}, {Monreal-Ibero}, {Verhamme},
  {Sandin}, {Steinmetz}, {Kollatschny}, {Krajnovi{\'c}}, {Kamann}, {Roth},
  {Erroz-Ferrer}, {Marino}, {Maseda}, {Wendt}, {Bacon}, {Dreizler}, {Richard},
  \& {Wisotzki}}]{2018A&A...611A..95W}
{Weilbacher}, P.~M., {Monreal-Ibero}, A., {Verhamme}, A., {et~al.} 2018, \aap,
  611, A95

\bibitem[{{Weilbacher} {et~al.}(2020){Weilbacher}, {Palsa}, {Streicher},
  {Bacon}, {Urrutia}, {Wisot zki}, {Conseil}, {Husemann}, {Jarno}, {Kelz},
  {P{\'e}contal-Rousset}, {Richard}, {Roth}, {Selman}, \&
  {Vernet}}]{musepipeline}
{Weilbacher}, P.~M., {Palsa}, R., {Streicher}, O., {et~al.} 2020, \aap, 641,
  A28

\bibitem[{{Xiao} {et~al.}(2018){Xiao}, {Stanway}, \&
  {Eldridge}}]{2018MNRAS.477..904X}
{Xiao}, L., {Stanway}, E.~R., \& {Eldridge}, J.~J. 2018, \mnras, 477, 904

\end{thebibliography}

\end{document}